\documentclass[aps,twocolumn,prd,superscriptaddress,showpacs,preprintnumbers,nofootinbib]{revtex4-1}

\usepackage{dsfont}
\usepackage{float,placeins}
\usepackage{graphicx}
\usepackage{amsmath,amssymb,amsthm,amsfonts,amsthm}
\usepackage{mathtools}
\usepackage{color}
\usepackage{hyperref}
\usepackage{siunitx}
\usepackage[caption=false]{subfig}

\def\eq#1{\eqref{#1}}
\def\Fig#1{Fig.~\ref{#1}}

\def\Sec#1{Sec.~\ref{#1}}
\def\App#1{App.~\ref{#1}}

\def\0#1#2{\frac{#1}{#2}}
\newcommand{\imag}{\text{i}}
\def\id{1\!\mbox{l}}
\newcommand{\pat}{k\partial_k}

\graphicspath{{./figures/}}

\newcommand{\gettitle}{Landau gauge Yang-Mills correlation functions}

\hypersetup{
  pdftitle={\gettitle},
  pdfauthor={Cyrol, Fister, Mitter, Pawlowski, Strodthoff},
  pdfkeywords={Yang-Mills theory} 
  	{correlations functions} 
  	{gluon mass gap} {confinement}  {propagators} 
  	{Running couplings} {functional renormalisation group},
  bookmarksopen=false,
  bookmarksnumbered=true
}

\begin{document}

\title{\gettitle}

\author{Anton K. Cyrol}
\affiliation{Institut f\"ur Theoretische
  Physik, Universit\"at Heidelberg, Philosophenweg 16, 69120
  Heidelberg, Germany} 

\author{Leonard Fister} \affiliation{Institut de Physique Th\'eorique,
  CEA Saclay, F-91191 Gif-sur-Yvette, France}

\author{Mario Mitter}
\affiliation{Institut f\"ur Theoretische
  Physik, Universit\"at Heidelberg, Philosophenweg 16, 69120
  Heidelberg, Germany} 

 \author{Jan M. Pawlowski} 
 \affiliation{Institut f\"ur Theoretische
  Physik, Universit\"at Heidelberg, Philosophenweg 16, 69120
  Heidelberg, Germany} 
\affiliation{ExtreMe Matter Institute EMMI, GSI, Planckstr. 1,
  D-64291 Darmstadt, Germany}

\author{Nils Strodthoff}
\affiliation{Institut f\"ur Theoretische
  Physik, Universit\"at Heidelberg, Philosophenweg 16, 69120
  Heidelberg, Germany} 
\affiliation{Nuclear Science Division, Lawrence Berkeley 
National Laboratory, Berkeley, CA 94720, USA}


\begin{abstract}
We investigate Landau gauge $SU(3)$ Yang-Mills theory in a
systematic vertex expansion scheme for the effective action with
the functional renormalisation group.
Particular focus is put on the dynamical creation of the gluon mass
gap at non-perturbative momenta and the consistent treatment of
quadratic divergences. The non-perturbative ghost and transverse
gluon propagators as well as the momentum-dependent ghost-gluon,
three-gluon and four-gluon vertices are calculated self-consistently
with the classical action as only input. The apparent convergence of
the expansion scheme is discussed and within the errors, our
numerical results are in quantitative agreement with available
lattice results.
\end{abstract}

\maketitle

\section{Introduction}

The past two decades have seen tremendous progress in the description
of QCD with functional approaches such as the
functional renormalisation group (FRG), Dyson-Schwinger equations
(DSE), and n-particle irreducible methods ($n$PI).  These approaches
constitute \emph{ab initio} descriptions of QCD in terms of quark and
gluon correlation functions. The full correlation functions satisfy a
hierarchy of loop equations that are derived from the functional FRG,
DSE and $n$PI relations for the respective generating functionals. By
now, systematic computational schemes are available, which can be
controlled by apparent convergence. In the present work on pure
Yang-Mills (YM) theory we complement the work in quenched QCD
\cite{Mitter:2014wpa}, where such a systematic expansion scheme has
been put forward within the FRG. Equipped with such a controlled
expansion, functional approaches to QCD are specifically interesting
at finite temperatures and large density, where reliable \emph{ab
  initio} theoretical predictions and experimental results are missing
at present.

Most progress with functional approaches has been made in Landau gauge
QCD, which has many convenient properties for non-perturbative
numerical computations. Applications of functional methods include the
first-ever calculation of qualitative non-perturbative Landau gauge
propagators as well as investigations of the phase structure of
QCD. For reviews see
\cite{Berges:2000ew,Roberts:2000aa,Alkofer:2000wg,Pawlowski:2005xe,
  Fischer:2006ub,Gies:2006wv,Schaefer:2006sr,Fischer:2008uz,Binosi:2009qm,%
Braun:2011pp,Maas:2011se,Sanchis-Alepuz:2015tha}, 
for applications to Yang-Mills theory see e.g.\
\cite{Ellwanger:1995qf,vonSmekal:1997ohs,Bergerhoff:1997cv,Gies:2002af,%
  Pawlowski:2003hq,Fischer:2003rp,Fischer:2004uk,Aguilar:2008xm,%
  Boucaud:2008ky,Tissier:2010ts,Quandt:2013wna,Quandt:2015aaa,Huber:2016tvc,Quandt:2016ykm}, 
and e.g.\ \cite{Feuchter:2004mk} for related studies in the Coulomb
gauge. The formal, algebraic, and numerical progress of the past
decades sets the stage for a systematic vertex expansion scheme of
Landau gauge QCD.  Quantitative reliability is then obtained with
apparent convergence \cite{Mitter:2014wpa} as well as further
systematic error controls inherent to the method, see e.g.\
\cite{Litim:2000ci,Litim:2001up,Pawlowski:2005xe,Schnoerr:2013bk,Pawlowski:2015mlf}.  In the aforementioned quenched QCD
investigation \cite{Mitter:2014wpa}, the gluon propagator was taken
from a separate FRG calculation in \cite{Fischer:2008uz,FPun}. This
gluon propagator shows quantitative agreement with the lattice
results, but has been obtained within an incomplete vertex expansion
scheme.  Therefore, the results \cite{Fischer:2008uz,FPun} for the YM
correlations functions give no access to systematic error estimates.
In general, many applications of functional methods to bound states
and the QCD phase diagram use such mixed approaches, where part of the
correlation functions is deduced from phenomenological constraints or
other external input. Despite the huge success of mixed approaches, a
full \emph{ab initio} method is wanted for some of the most pressing
open questions of strongly-interacting matter. The phase structure of
QCD at large density is dominated by fluctuations and even a partial
phenomenological parameter fixing at vanishing density is bound to
lead to large systematic errors \cite{Helmboldt:2014iya}. The same
applies to the details of the hadron spectrum, in particular with
regard to the physics of the higher resonances, which requires
knowledge about correlation functions deep in the complex plane.

In the present work we perform a systematic vertex expansion of the
effective action of Landau gauge YM theory within the functional
renormalisation group approach, discussed in \Sec{sec:setup}. The
current approximation is summarised in \Sec{sec:expsch}, which also
includes a comparison to approximations used in other works. This
\emph{ab initio} approach starts from the classical action. Therefore
the only parameter is the strong coupling constant $\alpha_s$ at a
large, perturbative momentum scale.  The most distinct feature of YM
theory is confinement, which is reflected by the creation of a gluon
mass gap in Landau gauge.  We discuss the necessity of consistent
infrared irregularities as well as mechanisms for the generation of a
mass gap in \Sec{sec:massgap}.  Numerical results from a
parameter-free, self-consistent calculation of propagators and
vertices are presented in \Sec{sec:mainresult}.  Particular focus is
put on the importance of an accurate renormalisation of the relevant
vertices.  We compare with corresponding DSE and lattice results and
discuss the apparent convergence of the vertex expansion.  Finally, we
present numerical evidence for the dynamic mass gap generation in our
calculation.  Further details, including a thorough discussion of the
necessary irregularities, can be found in the appendices.

\section{FRG flows for Yang-Mills theory in a vertex expansion}
\label{sec:setup}

Functional approaches to QCD and Yang-Mills theory are based on the
classical gauge fixed action of $SU(3)$ Yang-Mills theory.  In general
covariant gauges in four dimensions it is given by
\begin{align}
  S_{\rm cl}=\int_x\,\014 F_{\mu\nu}^a  F_{\mu\nu}^a   +\frac{1}{2\xi}\int_x\,(
  \partial_\mu A^a_\mu)^2 -\int_x\,\bar c^a \partial_\mu D^{ab}_\mu
  c^b\,.
\label{eq:sclassical} 
\end{align} 
Here, $\xi$ denotes the gauge fixing
parameter, which is taken to zero in the Landau gauge and $\int_x=\int \text{d}^4 x$.
The field
strength tensor and covariant derivative are given by
\begin{eqnarray}
  F^a_{\mu\nu} &=&  \partial_\mu A^a_\nu-\partial_\nu A^a_\mu+ 
  g f^{abc}A_\mu^b A_\nu^c\, ,\nonumber \\[2ex]
  D^{ab}_{\mu} & = &\delta^{ab}\partial_\mu-g f^{abc} A^c_\mu\,,
\end{eqnarray}
using the fundamental generators $T^a$, defined by 
\begin{eqnarray}
  \left[T^a,T^b\right] & =&   if^{abc}T^c\,,\qquad
  \text{tr}\left(T^aT^b\right) = \frac{1}{2}\delta^{ab}\,.
\end{eqnarray}
In general, our notation follows the one used in the works 
\cite{Mitter:2014wpa,Braun:2014ata,Rennecke:2015eba} of the fQCD
collaboration \cite{fQCD}.

\subsection{Functional Renormalisation Group}

We use the functional renormalisation group approach as a
non-perturbative tool to investigate Yang-Mills theory. The FRG is
built on a flow equation for the one-particle irreducible ($1$PI)
effective action or free energy of the theory, the Wetterich equation
\cite{Wetterich:1992yh}. It is based on Wilson's idea of introducing
an infrared momentum cutoff scale $k$. Here, this infrared
regularisation of the gluon and ghost fluctuations is achieved by
modifying the action $S_{\rm cl} \to S_{\rm cl}+\Delta S_k$ with
\begin{align}
\label{eq:dSk}
\Delta S_k=\int_x\,
\012  A_\mu^a\, R_{k,\mu\nu}^{ab} \, A_\nu^b  
+ \int_x\,\bar c^a\, R^{ab}_k\, c^b\,.
\end{align} 
The regulator functions $R_k$ are momentum-dependent masses that
suppress the corresponding fluctuations below momentum scales
$p^2\approx k^2$ and vanish in the ultraviolet for momenta $p^2\gg
k^2$. See \App{app:regulator} for details on the regulators used
in the present work.
Consequently, the effective action, $\Gamma_k[\phi]\,$, is
infrared regularised, where $\phi$ denotes the superfield
\begin{align}
\phi=(A, c,\bar c)\,.
\end{align}
The fluctuations of the theory are then successively taken into
account by integrating the flow equation for the effective action, see
e.g.\ \cite{Fister:2011uw,Fister:2013bh},
\begin{align}
{\partial_t}\Gamma_k[\phi] =  \int_p \,\012 \ G_{k,ab}^{\mu\nu}[\phi]
\,{\partial_t}R_{k,\mu\nu}^{ba} -\int_p
\,G^{ab}_k[\phi]\,{\partial_t}R^{ba}_k\,.
\label{eq:flow}
\end{align} 
where $\int_p=\int \text{d}^4p/(2 \pi)^4\,$. Here we have introduced the
RG-time $t=\ln (k/\Lambda)$ with a reference scale $\Lambda\,$, which is
typically chosen as the initial UV cutoff $\Lambda\,$.  Although this
flow equation comes in a simple one-loop form, it provides an exact
relation due to the presence of the full field-dependent propagator,
\begin{align}
  G_k[\phi](p,q)= \01{\Gamma_k^{(2)}[\phi]+R_k }(p,q) \,,
\end{align} 
on its right-hand side. Furthermore, the flow is infrared and
ultraviolet finite by construction. Via the integration of momentum
shells in the Wilsonian sense, it connects the ultraviolet, bare
action $S_{\rm cl}=\Gamma_{k\rightarrow\Lambda\rightarrow\infty}$ with
the full quantum effective action $\Gamma=\Gamma_{k\rightarrow0}\,$.

The flow equations for propagators and vertices are obtained by taking
functional derivatives of \eq{eq:flow}.  At the vacuum expectation
values, these derivatives give equations for the $1$PI correlation
functions $\Gamma_k^{(n)}=\delta^n\Gamma_k/\delta\phi^n\,$, which
inherit the one-loop structure of \eq{eq:flow}. As the
cutoff-derivative of the regulator functions, $\partial_t R_k$, decays 
sufficiently fast for large momenta, the momentum integration in
\eq{eq:flow} effectively receives only contributions for momenta
$p^2\lesssim k^2\,$. Furthermore, the flow depends solely on dressed
vertices and propagators, leading to a consistent RG and momentum
scaling for each diagram resulting from derivatives of \eq{eq:flow}.
Despite its simple structure, the resulting system of equations does
not close at a finite number of correlation functions. In general,
higher derivatives up to the order $\Gamma_k^{(n+2)}$ of the effective
action appear on the right hand side of the functional relations for
the correlation functions $\Gamma_k^{(n)}\,$.

\subsection{Vertex expansion of the effective action}
\label{sec:expsch}
\begin{figure}
	\centering
	\includegraphics[width=0.48\textwidth]{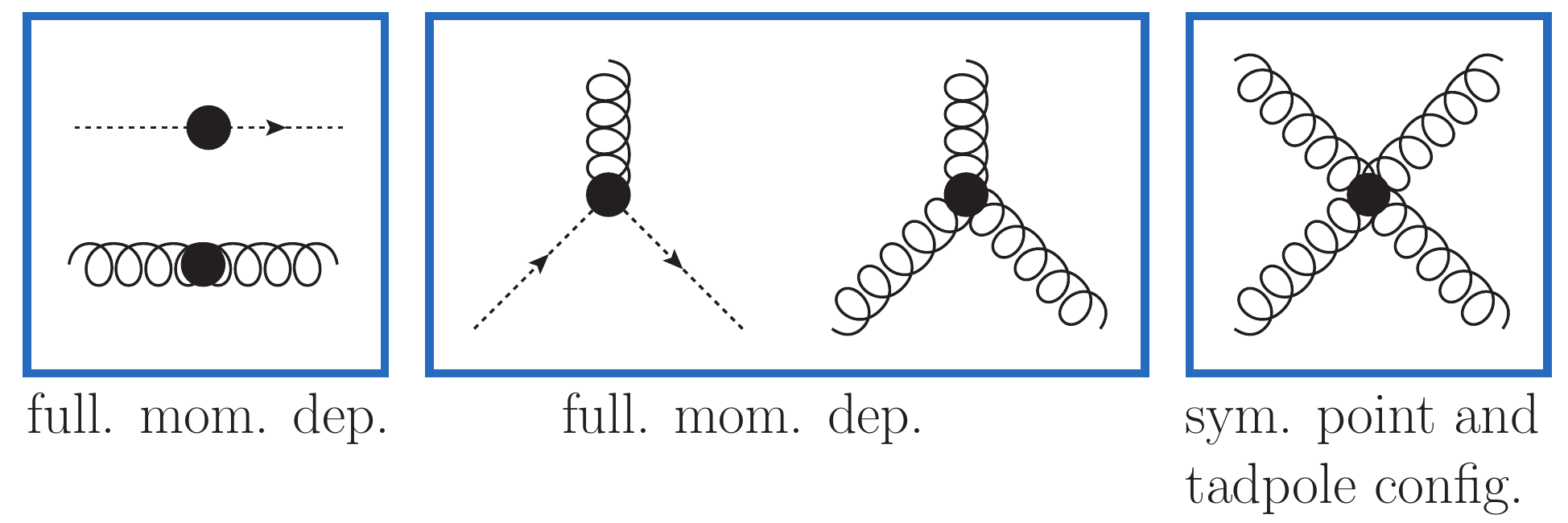}
	\caption{ Approximation for the effective action.  Only
          classical tensor structures are included.  See
          \Fig{fig:diagrams} for diagrams that contribute to the
          individual propagators and vertices.  }
	\label{fig:eff_action}
\end{figure}
\begin{figure}
  	\hspace*{-0.1cm}\includegraphics[width=0.49\textwidth]{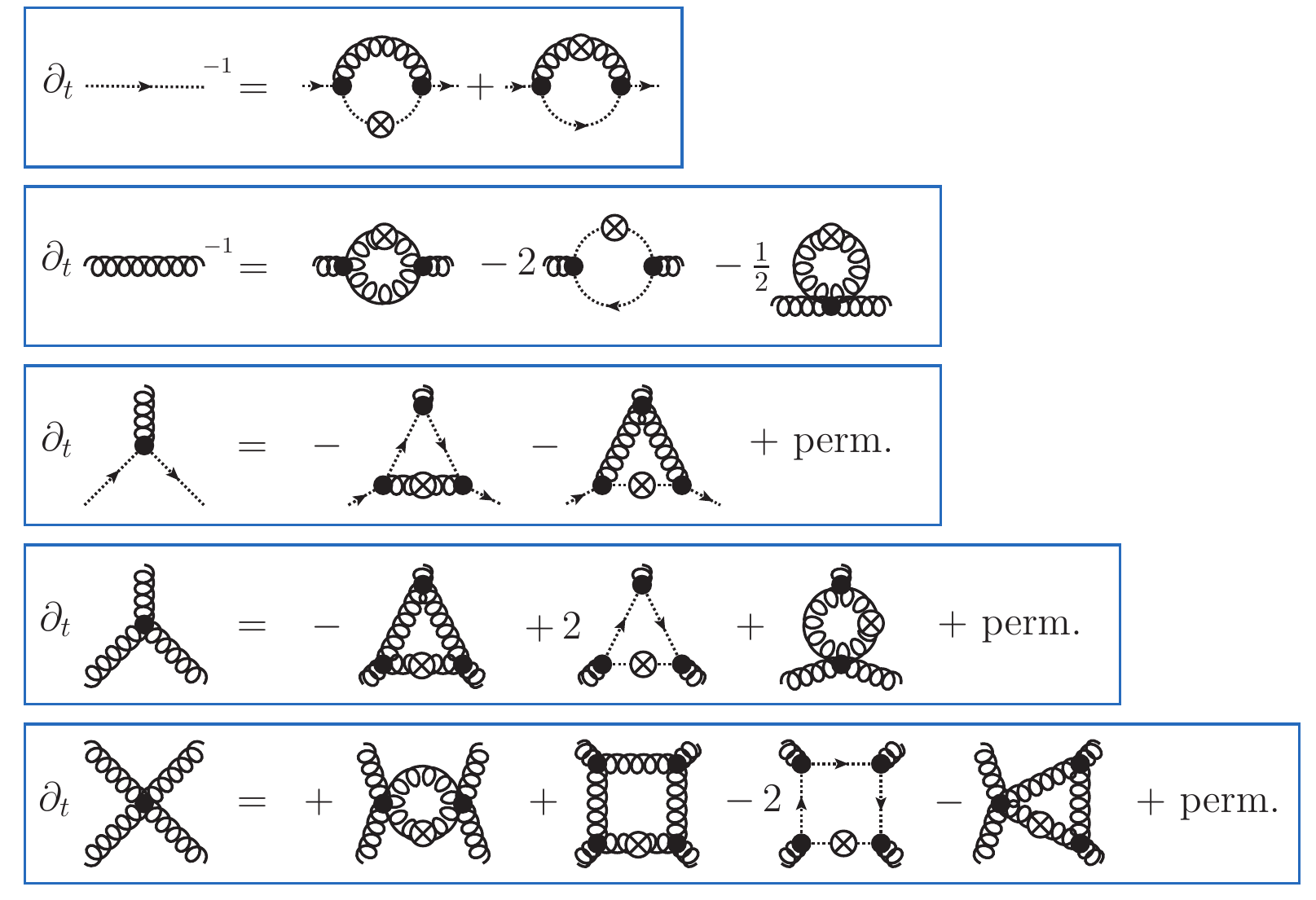}
        \caption{Diagrams that contribute to the truncated flow of
          propagators and vertices. Wiggly (dotted) lines correspond
          to dressed gluon (ghost) propagators, filled circles denote
          dressed ($1$PI) vertices and regulator insertions are
          represented by crossed circles.  Distinct permutations
          include not only (anti-)symmetric permutations of external
          legs but also permutations of the regulator insertions.  }
	\label{fig:diagrams}
\end{figure}

The structural form of the functional equations discussed in the
previous section necessitates the use of approximations in most
practical application. One systematic expansion scheme is the vertex
expansion, i.e.\ an expansion of the effective action in terms of
$1$PI Green's functions. This yields an infinite tower of coupled
equations for the correlation functions that has to be truncated at a
finite order. This expansion scheme allows a systematic error estimate
in terms of apparent convergence upon increasing the expansion order
or improving further approximations for example in the momentum
resolution or tensor structures of the included correlation functions.
We discuss the convergence of the vertex expansion in
\Sec{sec:truncationcheck}.

Here we calculate the effective action of $SU(3)$ Yang-Mills theory in
Landau gauge within a vertex expansion, see \Fig{fig:eff_action} for a
pictorial representation.  The diagrams contributing to the resulting
equations of the constituents of our vertex expansion are summarised
graphically in \Fig{fig:diagrams}.  The lowest order contributions
in this expansion are the inverse gluon and ghost propagators
parameterised via
\begin{align}
  [\Gamma^{(2)}_{AA}]^{ab}_{\mu\nu}(p) &= Z_A(p)\, p^2\, \delta^{ab}\,
  \Pi^{\bot}_{\mu\nu}(p) +\01\xi \delta^{ab} p_\mu p_\nu\ ,
  \nonumber\\[2ex]
  [\Gamma^{(2)}_{\bar c c}]^{ab}(p) &= Z_c(p)\, p^2\, \delta^{ab} \,,
\label{eq:propagators}
\end{align}
with dimensionless scalar dressing functions $1/Z_A$ and $1/Z_c$. Here,
$\Pi^{\bot}_{\mu\nu}(p)=\delta_{\mu\nu}-p_\mu p_\nu/p^2$ denotes the
corresponding projection operator.
We use this
splitting in tensor structures with canonical momentum dimension and
dimensionless dressings also for the higher order vertices.

On the three-point level we include the full transverse ghost-gluon
vertex and the classical tensor structure of the three-gluon vertex
\begin{align}
  [\Gamma^{(3)}_{A \bar c c}]^{abc}_\mu (p,q) &= Z_{A \bar c c,\bot}(|p|,|q|,t)
  [\mathcal{T}_{A \bar c c,{\rm cl}}]^{abc}_{\mu}(p,q)\,,\nonumber\\[2ex]
  [\Gamma^{(3)}_{A^3}]^{abc}_{\mu\nu\rho} (p,q) &= Z_{A^3,\bot}(|p|,|q|,t)
  [\mathcal{T}_{A^3,{\rm cl}}]^{abc}_{\mu\nu\rho}(p,q)\,.
\label{eq:threepoint}
\end{align}
Here, the momentum $p$ $(q)$ corresponds to the indices $a$ $(b)$ and
$t$ denotes the cosine of the angle between the momenta $p$ and $q\,$.
The classical tensor structure of the vertices has been summarised as
$\mathcal{T}_{A^3,{\rm cl}}$ and $\mathcal{T}_{A \bar c c,{\rm cl}}\,$,
which are listed explicitly in \App{app:tensorstructures}.  In the
case of the transversally projected ghost-gluon vertex,
$\mathcal{T}_{A \bar c c,{\rm cl}}$ represents already a full basis
whereas a full basis for the transversally projected three-gluon
vertex consists of four elements. However, the effect of non-classical
tensor structures has been found to be subleading in this case
\cite{Eichmann:2014xya}.

The most important four-point function is given by the four-gluon
vertex, which appears already on the classical level. Similarly to the
three-gluon vertex, we approximate it with its classical tensor
structure
\begin{align}
  [\Gamma_{A^4}^{(4)}]^{abcd}_{\mu\nu\rho\sigma}(p,q,r) &=
  Z_{A^4,\bot}(\bar{p})
  [\mathcal{T}_{A^4,{\rm cl}}]^{abcd}_{\mu\nu\rho\sigma}(p,q,r)\,,
\label{eq:fourgluon}
\end{align}
see \App{app:tensorstructures} for details.  The dressing function of
the four-gluon vertex is approximated from its momentum dependence at
the symmetric point via the average momentum $\bar{p}\equiv
\sqrt{p^2+q^2+r^2+(p+q+r)^2}/2\,$, which has been shown to be a good
approximation of the full momentum dependence
\cite{Cyrol:2014kca,Cyrol:2014mt}.  To improve this approximation
further, we additionally calculate the momentum dependence of the
four-gluon dressing function $Z_{A^4,\bot}(|p|,|q|,t)$ on the special
configuration $(p,q,r)=(p,q,-p)\,$. We use this special configuration
exclusively in the tadpole diagram of the gluon propagator equation,
cf.\ \Sec{sec:truncationcheck}.  We show the difference between the
special configuration and the symmetric point approximation in the
appendix in \Fig{fig:fourGluonTadpoleDressing}.  Although the
four-gluon vertex has been the subject of several studies
\cite{Kellermann:2008iw,Binosi:2014kka,Gracey:2014ola,Cyrol:2014kca,Cyrol:2014mt},
no fully conclusive statements about the importance of additional
non-classical tensors structures are available.

In summary we have taken into account the propagators and the fully
momentum-dependent classical tensor structures of the three-point
functions, as well as selected momentum-configurations of the gluon
four-point function, see the paragraph above, and
\App{app:tensorstructures}. For a comparison of the current approximation with that 
used in other functional works one has to keep in mind that FRG,
Dyson-Schwinger or $n$PI equations implement different resummation
schemes. Thus, even on an identical approximation level of a
systematic vertex expansion, the included resummations differ. 
 
In the present work we solve the coupled system of all
momentum-dependent classical vertex structures and propagators. In former 
works with functional methods, see e.g.\ 
\cite{Ellwanger:1995qf,vonSmekal:1997ohs,Bergerhoff:1997cv,Gies:2002af,%
  Pawlowski:2003hq,Fischer:2003rp,Fischer:2004uk,Kellermann:2008iw,Aguilar:2008xm,%
  Boucaud:2008ky,Tissier:2010ts,Huber:2012kd,Aguilar:2013xqa,Pelaez:2013cpa,Blum:2014gna,Eichmann:2014xya,%
  Gracey:2014mpa,Gracey:2014ola,Huber:2014isa,Williams:2015cvx,Binosi:2014kka,Cyrol:2014kca,Cyrol:2014mt}, 
only subsets of these correlation functions have
been coupled back.  A notable exception is \cite{Huber:2016tvc}, where
a similar self-consistent approximation has been used for three-dimensional
Yang-Mills theory.

\subsection{Modified Slavnov-Taylor identities and transversality in
  Landau gauge}
\label{sec:mSTIandVert_sub}

In Landau gauge, the dynamical system of correlation functions
consists only of the transversally projected correlators
\cite{Fischer:2008uz}.  Those with at least one longitudinal gluon leg
do not feed back into the dynamics.  To make these statements precise,
it is useful to split correlation functions into purely transverse
components and their complement with at least one longitudinal gluon
leg. The purely transverse vertices $\Gamma^{(n)}_{\bot}\,$, are defined
by attaching transverse projection operators to the corresponding
gluon legs,
\begin{align}
\label{eq:purely_transverse} 
  &\left[\Gamma^{(n)}_{\bot}\right]_{\mu_1\cdots\mu_{n_A}} \equiv
  \Pi^{\bot}_{\mu_1\nu_1} \cdots \Pi^{\bot}_{\mu_{n_A}\nu_{n_A}}
  \left[\Gamma^{(n)}\right]_{\nu_1\cdots\nu_{n_A}}\,,
\end{align}
where $n_A$ is the number of gluon legs and group indices and momentum
arguments have been suppressed for the sake of brevity.  This defines
a unique decomposition of $n$-point functions into
\begin{align}
\label{eq:longitudinal} 
\Gamma^{(n)}
=\Gamma^{(n)}_{\bot}+\Gamma^{(n)}_{\text{L}}\,,
\end{align}
where the longitudinal vertices $\Gamma^{(n)}_{\text{L}}$, have at
least one longitudinal gluon leg.  Consequently, they are always
projected to zero by the purely transverse projection operators of
\eq{eq:purely_transverse}.

Functional equations for the transverse correlation functions close in
the Landau gauge, leading to the structure \cite{Fischer:2008uz},
\begin{align}\label{eq:closedFun} 
\Gamma^{(n)}_{\bot}
={\rm Diag}[\{\Gamma^{(n)}_{\bot}\}]\,.
\end{align} 
In \eq{eq:closedFun} Diag stands for diagrammatic expressions of
either integrated FRG, Dyson-Schwinger or $n$PI equations.  Equation
\eq{eq:closedFun} follows from the fact that all internal legs are
transversally projected by the Landau gauge gluon propagator. Hence,
by using transverse projections for the external legs one obtains
\eq{eq:closedFun}.  In contradistinction to this, the functional
equations for the vertices with at least one longitudinal gluon leg,
$\Gamma_{\rm L}^{(n)}$, are of the form
\begin{align}
\label{eq:closedFunL} 
\Gamma^{(n)}_{\rm L}
={\rm Diag}[\{\Gamma^{(n)}_{\rm L}\}, \{\Gamma^{(n)}_{\bot}\}]\,.
\end{align} 
In other words, the solution of the functional equations
\eq{eq:closedFunL} for $\Gamma^{(n)}_{\rm L}$ requires also the
solution of the transverse set of equations \eq{eq:closedFun}.

In the present setting, gauge invariance is encoded in modified
Slavnov-Taylor identities (mSTIs) and Ward-Takahashi identities
(mWTIs). They are derived from the standard Slavnov-Taylor identities
(STIs) by including the gauge or BRST variations of the regulator
terms, see
\cite{Ellwanger:1994iz,Ellwanger:1995qf,D'Attanasio:1996jd,Igarashi:2001mf,
  Pawlowski:2005xe,Igarashi:2016gcf} for details. The mSTIs are of the
schematic form
\begin{align}
\label{eq:longFun}
  \Gamma_{\rm L}^{(n)}&={\rm mSTI} [\{\Gamma_{\rm L}^{(n)}\}\,,\,
  \{\Gamma_{\bot}^{(n)}\}\,,\,R_k]\,,		      
\end{align}
which reduce to the standard STIs in the limit of vanishing regulator,
$R_k\equiv 0$. The STIs and mSTIs have a similar structure as
\eq{eq:closedFunL} and can be used to obtain information about the
longitudinal part of the correlators.  Alternatively, they provide a
non-trivial consistency check for approximate solutions of
\eq{eq:closedFunL}.

\subsubsection*{Consequences of the STIs \& mSTIs}
\label{sec:STImSTI}

For the purposes of this work, the most important effect of the
modification of the STIs due to the regulator term is that it leads to
a non-vanishing gluon mass parameter \cite{Ellwanger:1994iz},
\begin{align} 
  \Delta_{\rm mSTI} \left[
    \Gamma_{AA}^{(2)}\right]_{\mu\nu}^{ab}\varpropto \delta^{ab}\,
  \delta_{\mu\nu}\, \alpha(k)\, k^2\,.
  \label{eq:mSTI_mass}
\end{align}
At $k=0$, where the regulators vanish, this modification disappears,
as the mSTIs reduce to the standard STIs. In particular, this entails
that, at $k=0$, the inverse longitudinal gluon propagator,
$\Gamma_{AA,\rm L}^{(2)}$, reduces to the classical one, solely
determined by the gauge fixing term
\begin{align}
\label{eq:LSTI}
  p_\mu \left( [{\Gamma_{AA,\rm L}^{(2)}}]_{\mu\nu}^{ab}(p) - [{S_{AA,\rm
        L}^{(2)}}]_{\mu\nu}^{ab}(p)\right) = 0\,.
\end{align}
This provides a unique condition for
determining the value of the gluon mass parameter \eq{eq:mSTI_mass}
at the ultraviolet initial scale $\Lambda$. However, it can
only serve its purpose, if the longitudinal system
is additionally solved.

One further conclusion from \eq{eq:longFun} is that the mSTIs do not
constrain the transverse correlation functions without further
input. This fact is not in tension with one of the main applications
of STIs in perturbation theory, i.e.\ relating the running of the
relevant vertices of Yang-Mills theory that require renormalisation.
As Yang-Mills theory is renormalisable, only the classical vertex
structures are renormalised and hence the renormalisation functions of
their transverse and longitudinal parts have to be identical. 

As an instructive example we consider the ghost-gluon vertex. For this
example and the following discussions we evaluate the STIs within the
approximation used in the present work: on the right hand side of the
STIs we only consider contributions from the primitively divergent
vertices.  In particular, this excludes contributions from the
two-ghost--two-gluon vertex. The ghost-gluon vertex can be
parameterised with two tensor structures,
\begin{align}
  [\Gamma^{(3)}_{A\bar c c}]_{\mu}^{abc}(p,q) = \imag f^{abc}\Bigl[
  q_\mu Z_{A \bar cc,\rm cl}(p,q) + p_\mu Z_{A\bar c c,\rm ncl}
  (p,q)\Bigr]\,.
\label{eq:classquantsplit}
\end{align}
In \eq{eq:classquantsplit} we have introduced two dressing functions
$Z_{A \bar cc, \rm cl}$ and $Z_{A \bar cc,\rm ncl}$ as functions of
the gluon momentum $p$ and anti-ghost momentum $q\,$.  In a general
covariant gauge only $Z_{A \bar cc,\rm cl}$ requires
renormalisation. Similar splittings into a classical tensor structure
and the rest can be used in other vertices. Trivially, this property 
relates the perturbative RG-running of the transverse and longitudinal
projections of the classical tensor structures. Then, the STIs can be
used to determine the perturbative RG-running of the classical tensor
structures, leading to the well-known perturbative relations
\begin{equation}
  \frac{Z_{A\bar c c, \rm cl}^2}{Z_c^2 Z_A}=
  \frac{Z^2_{A^3,\rm cl}}{Z_A^3}=\frac{Z_{A^4,\rm cl}}{Z_A^2}\,, 
\label{eq:RGrel}
\end{equation}
at the renormalisation scale $\mu$. Consequently, \eq{eq:RGrel} allows
for the definition of a unique renormalised two-loop coupling
$\alpha_s(\mu)$ from the vertices.

The momentum dependent STIs can also be used to make the relation
\eq{eq:RGrel} momentum-dependent. Keeping only the classical tensor
structures, we are led to the momentum dependent running couplings
\begin{align}
  \alpha_{A\bar c c}(p) &= \0{1}{4 \pi}\,\frac{Z_{A\bar cc,\bot }^2(p)}
  { Z_A(p)\,Z_c^2(p)}\,,\nonumber\\[2ex]
  \alpha_{A^3}(p) &= \0{1}{4 \pi} \,\frac{Z_{A^3,\bot }^2(p)}
  {Z_A^3(p)}\ ,\nonumber\\[2ex]
  \alpha_{A^4}(p) &= \0{1}{4 \pi}\, \frac{Z_{A^4,\bot }(p)}{
    Z_A^2(p)}\,, 
\label{eq:runcoup}
\end{align}
where the used transverse projection is indicated by the subscript
$\bot$, for details see \App{app:tensorstructures}.  Additionally, the
vertices appearing in \eq{eq:runcoup} are evaluated at the symmetric
point, see \Sec{sec:truncationcheck} for the precise definition.  The
STIs and two-loop universality demand that these running couplings
become degenerate at large perturbative momentum scales, where the
longitudinal and transverse parts of the vertices agree.

In Landau gauge, the ghost-gluon vertex is not renormalised on
specific momentum configurations, and we can alternatively define a
running coupling from the wave function renormalisation of ghost and
gluon \cite{vonSmekal:1997ohs,vonSmekal:2009ae},
\begin{align}
\label{eq:propcoupling}
\alpha_s(p)= \0{1}{4\pi} \,\0{g^2 }{Z_A(p) Z_c^2(p)}\,.
\end{align}
Note that the momentum-dependence of the running coupling
\eq{eq:propcoupling} does not coincide with that of the corresponding
running couplings obtained from other vertices, i.e.\ \eq{eq:runcoup}.
This is best seen in the ratio $\alpha_{A\bar c c}(p)/\alpha_s(p)=
Z_{A\bar c c,\bot}^{2}(p)/g^2\,$. In this context we also report on an important
result for the quark-gluon vertex coupling,
\begin{align}
\label{eq:quarkgluon}
  \alpha_{A\bar q q}(p)=\0{1}{4\pi} \,\0{Z_{A\bar q q,\bot}(p)^2 }{Z_A(p)
  Z_q(p)^2}\,,
\end{align}
with the dressing function of the classical tensor structure of the
quark-gluon vertex $Z_{A\bar q q,\bot}(p)^2\,$ 
and the quark dressing function $1/Z_q(p)$~\cite{Mitter:2014wpa}. The solution of the
corresponding STI reveals that the quark-gluon vertex coupling
$\alpha_{A\bar q q}$ agrees perturbatively with $\alpha_s(p)$ in
\eq{eq:propcoupling}, and hence it differs from the other vertex
couplings in \eq{eq:runcoup}. Note that the present truncation only
considers contributions from primitively divergent
vertices. Accordingly, the two-quark--two-ghost vertex contribution in
the STI for the quark-gluon vertex, see e.g.\ \cite{Alkofer:2000wg}, has
been dropped.

\section{Confinement, gluon mass gap, \newline and irregularities}
\label{sec:massgap}

It has been shown in
\cite{Braun:2007bx,Marhauser:2008fz,Braun:2009gm,Braun:2010cy,Fister:2013bh} that a
mass gap in the gluon propagator signals confinement in QCD in
covariant gauges. Furthermore, in Yang-Mills theory formulated in
covariant gauges, the gapping of the gluon relative to the ghost is
necessary and sufficient for producing a confining potential for the
corresponding order parameter, the Polyakov loop. Hence, understanding
the details of the dynamical generation of a gluon mass gap gives
insight into the confinement mechanism.

This relation holds for all potential infrared closures of the
perturbative Landau gauge. The standard infrared closure 
corresponds to a full average over all Gribov regions. This leads to
the standard Zinn-Justin equation as used in the literature, e.g.\
\cite{Alkofer:2000wg}. In turn, the restriction to the first Gribov
regime can be implemented within the refined Gribov-Zwanziger
formalism, e.g.\
\cite{Dudal:2007cw,Dudal:2008sp,Dudal:2009xh,Dudal:2011gd,Capri:2015ixa},
that leads to infrared modifications of the STIs. In the following we
discuss the consequences of the standard STIs, a discussion of the
refined Gribov-Zwanziger formalism is deferred to future work.
\begin{figure*}
	\centering
	\includegraphics[width=0.48\textwidth]{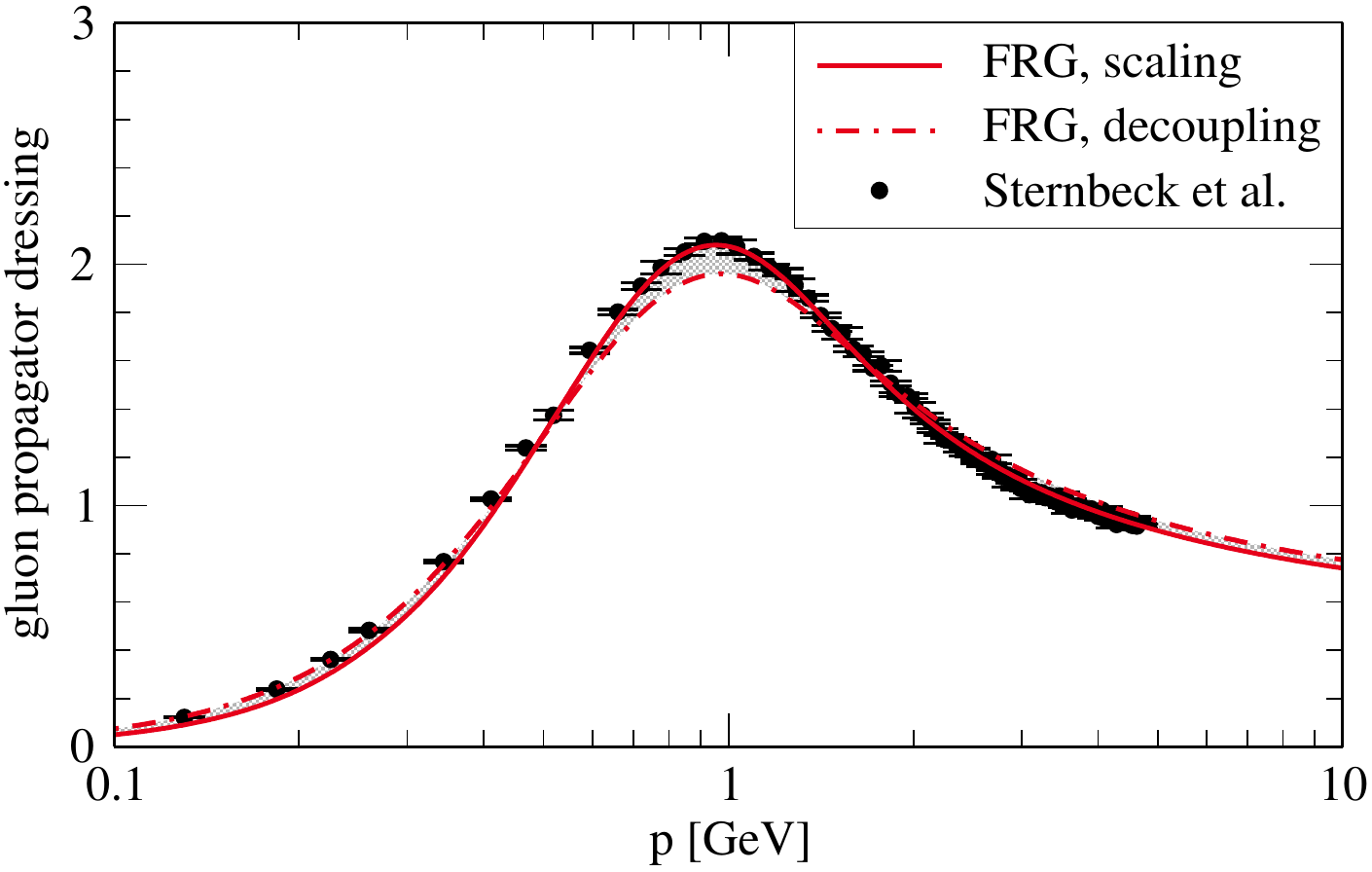}
        \hfill
	\includegraphics[width=0.489\textwidth]{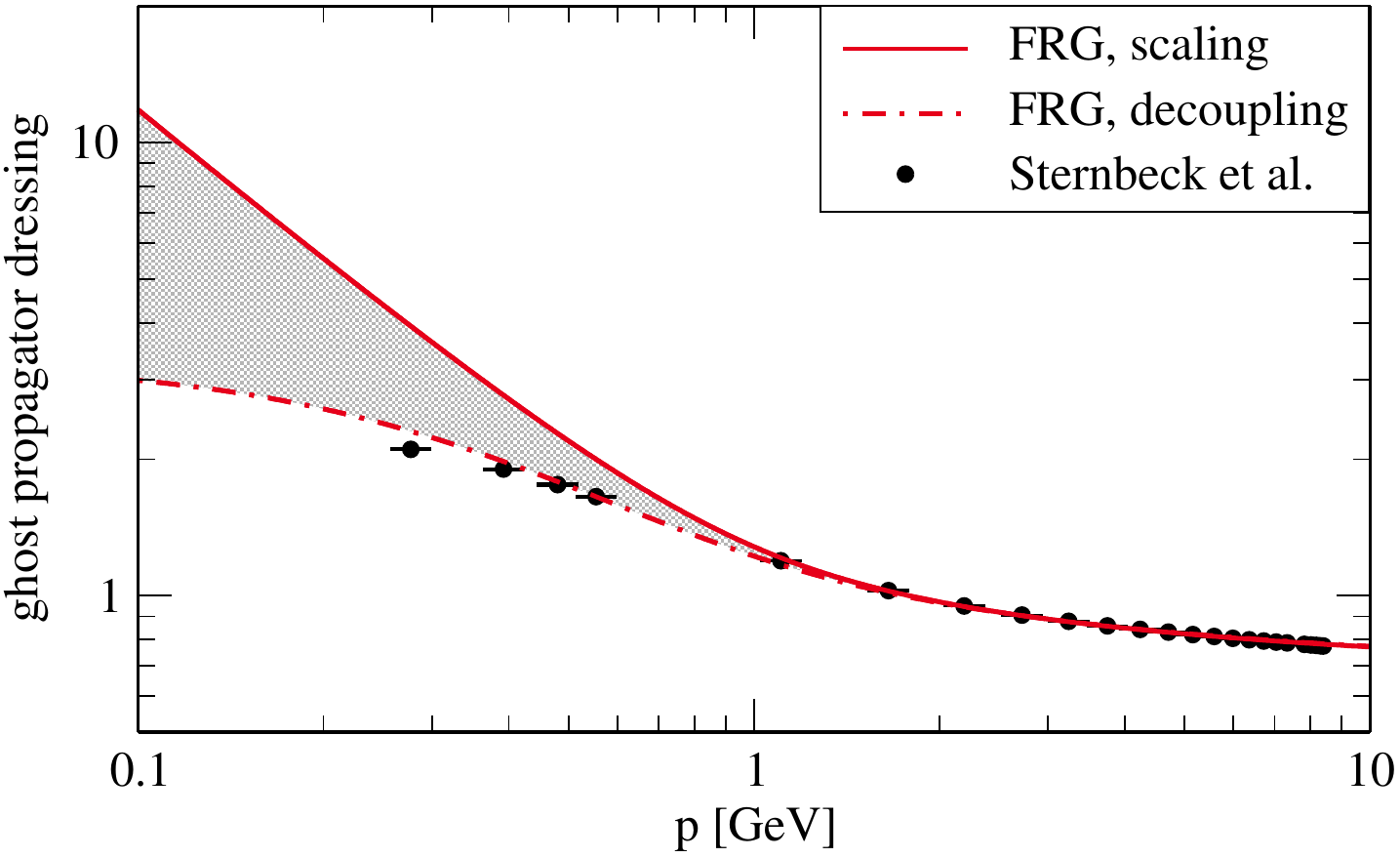}
	\caption{Gluon dressing $1/Z_A$ (left) and ghost dressing $1/Z_c$
	(right) in comparison to the lattice results
	from \cite{Sternbeck:2006cg}. The scale setting and normalisation procedures are described in \App{app:rescaling}.}
	  \label{fig:main_result}
\end{figure*}

\subsection{Gluon mass gap and irregularities}
\label{sec:gluonmassirregularities}

In order to study the dynamical generation of the mass gap, we first
discuss the consequences of the STI for the longitudinal gluon two
point function \eq{eq:LSTI}. It states that no quantum fluctuations
contribute to the inverse longitudinal gluon propagator, i.e.\ the
longitudinal gluon propagator is defined by the gauge fixing
term. Therefore, the dynamical creation of a gluon mass gap requires
different diagrammatic contributions to the longitudinal and
transverse gluon mass parameter. The discussion of the prerequisites
for meeting this condition is qualitatively different for the scaling
and the decoupling solutions.  Hence, these two cases are discussed
separately.

The scaling solution is characterised by the infrared behaviour
\cite{vonSmekal:1997ohs,Zwanziger:2001kw,Lerche:2002ep,Fischer:2002eq,%
Pawlowski:2003hq,Alkofer:2004it,Fischer:2006vf,Alkofer:2008jy,Fischer:2009tn}
\begin{align}\nonumber 
  \lim\limits_{p\rightarrow 0}Z_c(p^2)&\varpropto (p^2)^{\kappa}\,,\\[2ex]
  \lim\limits_{p\rightarrow 0}Z_A(p^2)&\varpropto (p^2)^{-2\,
    \kappa}\,,
\label{eq:scaling_sol}
\end{align}
with the scaling coefficient $1/2<\kappa<1$.  A simple calculation
presented in \App{app:irregularities} shows that the ghost loop with
an infrared constant ghost-gluon vertex and scaling ghost propagator
is already capable of inducing a splitting in the longitudinal and
transverse gluon mass parameter.

Next we investigate the decoupling solution, e.g.\
\cite{Aguilar:2008xm,Boucaud:2008ky}, which scales with
\begin{align}\nonumber
 \lim\limits_{p\rightarrow 0}Z_c(p^2)&\varpropto 1\, ,\\[2ex]
 \lim\limits_{p\rightarrow 0}Z_A(p^2)&\varpropto (p^2)^{-1}\,,
 \label{eq:decoupling_sol}
\end{align}
at small momenta. Assuming vertices that are regular in the limit of
one vanishing gluon momentum, one finds that all diagrammatic
contributions to the longitudinal and transverse gluon mass parameter
are identical.  For example, if the ghost loop were to yield a 
non-vanishing contribution to the gluon mass gap, the ghost-gluon
vertex would have to be a function of the angle $\theta=\arccos(t)$
between the gluon and anti-ghost momenta $p$ and $q$,
\begin{align}
\label{eq:irregular}
  \lim\limits_{|p|\rightarrow 0}[\Gamma^{(3)}_{A\bar c
    c}{}]_{\mu}^{abc}(|p|,|q|,t) = [\Gamma^{(3)}_{A\bar c
    c}{}]_{\mu}^{abc}(0,|q|,t)\,,
\end{align}
even in the limit of vanishing gluon momentum $|p|\rightarrow
0\,$. Since the above limit depends on the angle, the vertex is
irregular. See \App{app:irregularities} for more details on this
particular case. Similar conclusions can be drawn for all vertices
appearing in the gluon propagator equation.  Consequently, if all
vertices were regular, no gluon mass gap would be created.  In
particular, regular vertices would entail the absence of
confinement. The necessity of irregularities for the creation of a
gluon mass gap was already realised by Cornwall~\cite{Cornwall:1981zr}.

In the light of these findings it is interesting to investigate the
consistency of irregularities with further Slavnov-Taylor
identities. Therefore, we consider the Slavnov-Taylor identity of the
three-gluon vertex, e.g.\ \cite{Alkofer:2000wg},
\begin{widetext}
  \begin{align}
  \label{eq:STIclass3gl} 
  \imag r_\rho
    [\Gamma^{(3)}_{A^3}]{}^{abc}_{\mu\nu\rho}(p,q,r)\propto f^{abc}
    \0{1}{Z_c(r^2)}\left( \tilde G_{\mu\sigma}(p,q) q^2
      Z_A(q^2)\Pi^\bot_{\sigma\nu}(q) -\tilde G_{\nu\sigma}(q,p)
      p^2 Z_A(p^2)\Pi^\bot_{\sigma\mu}(p) \right)\,,
\end{align}
\end{widetext}
where $\tilde G_{\mu \nu}$ relates to the ghost-gluon vertex via
\begin{align}
\label{eq:STIGtilde}
  [\Gamma^{(3)}_{A \bar c c} ]_{\mu}^{abc}(p,q)=\imag g f^{abc} q_\nu
  \tilde G_{\mu\nu}(p,q)\,.
\end{align}
For a regular $\tilde G_{\mu\nu}$ in the limit $p\rightarrow 0$ 
in \eq{eq:STIclass3gl}, the
scaling solution leads to a singular contribution of the type
\begin{align}\label{eq:3g_sing_scaling}
  \lim\limits_{p\rightarrow 0} (p^2)^{1-2\kappa}\, \tilde
  G_{\nu\sigma}(q,0)\, \Pi^\bot_{\sigma\mu}(p) + \rm regular\,,
\end{align}
where $\kappa$ is the scaling coefficient from
\eq{eq:scaling_sol}. This is consistent with the expected scaling
exponent of the three-gluon vertex in this limit
\cite{Alkofer:2008jy}. In the same limit, the decoupling solution
leads to a singular contribution of the form
\begin{align}
\label{eq:3g_sing_decoupling}
  \lim\limits_{p\rightarrow 0} \tilde G_{\nu\sigma}(q,0)\,
  \Pi^\bot_{\sigma\mu}(p) + \rm regular\,.
\end{align}
Since the transverse projector $\Pi^\bot_{\sigma\mu}(p)$ introduces an
angular dependence in the limit $p\rightarrow 0\,$, the STI again
demands an irregularity in limit of one vanishing momentum.  Note that
this is just a statement about the three-gluon vertex projected with
one non-zero longitudinal leg $r_\rho\,$.  Although this momentum
configuration does not enter the gluon mass gap directly, crossing
symmetry implies the necessary irregularity. In summary, these
arguments illustrate that also the three-gluon vertex STI is
consistent with the necessity of irregularities for both types of
solutions.

We close the discussion of vertex irregularities with the remark that
the infrared modification of the propagator-STI in the refined 
Gribov-Zwanziger formalism may remove the necessity for irregularities
in the vertices.
\begin{figure*}
	\centering
	\includegraphics[width=0.48\textwidth]{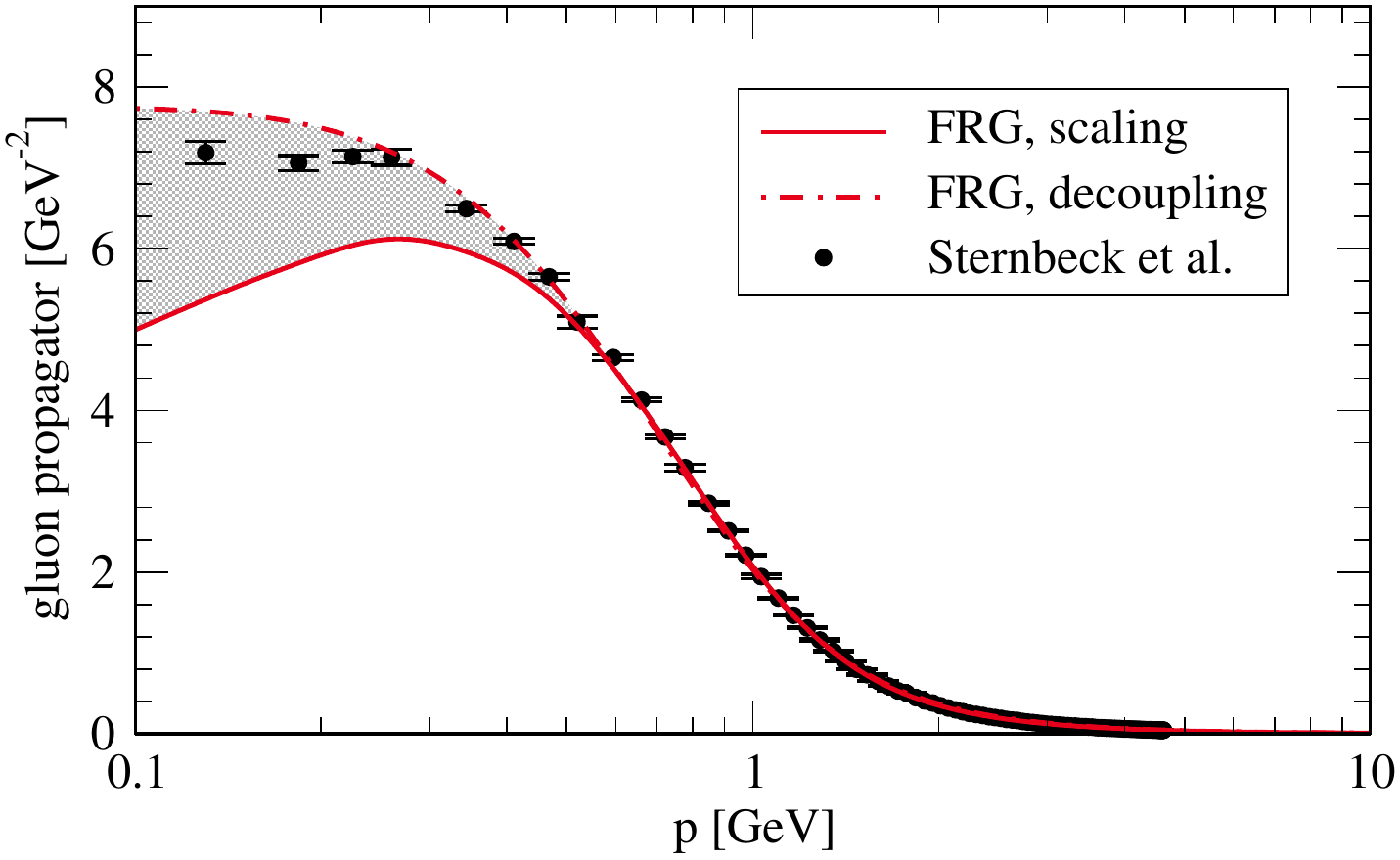}
\hfill
	\includegraphics[width=0.496\textwidth]{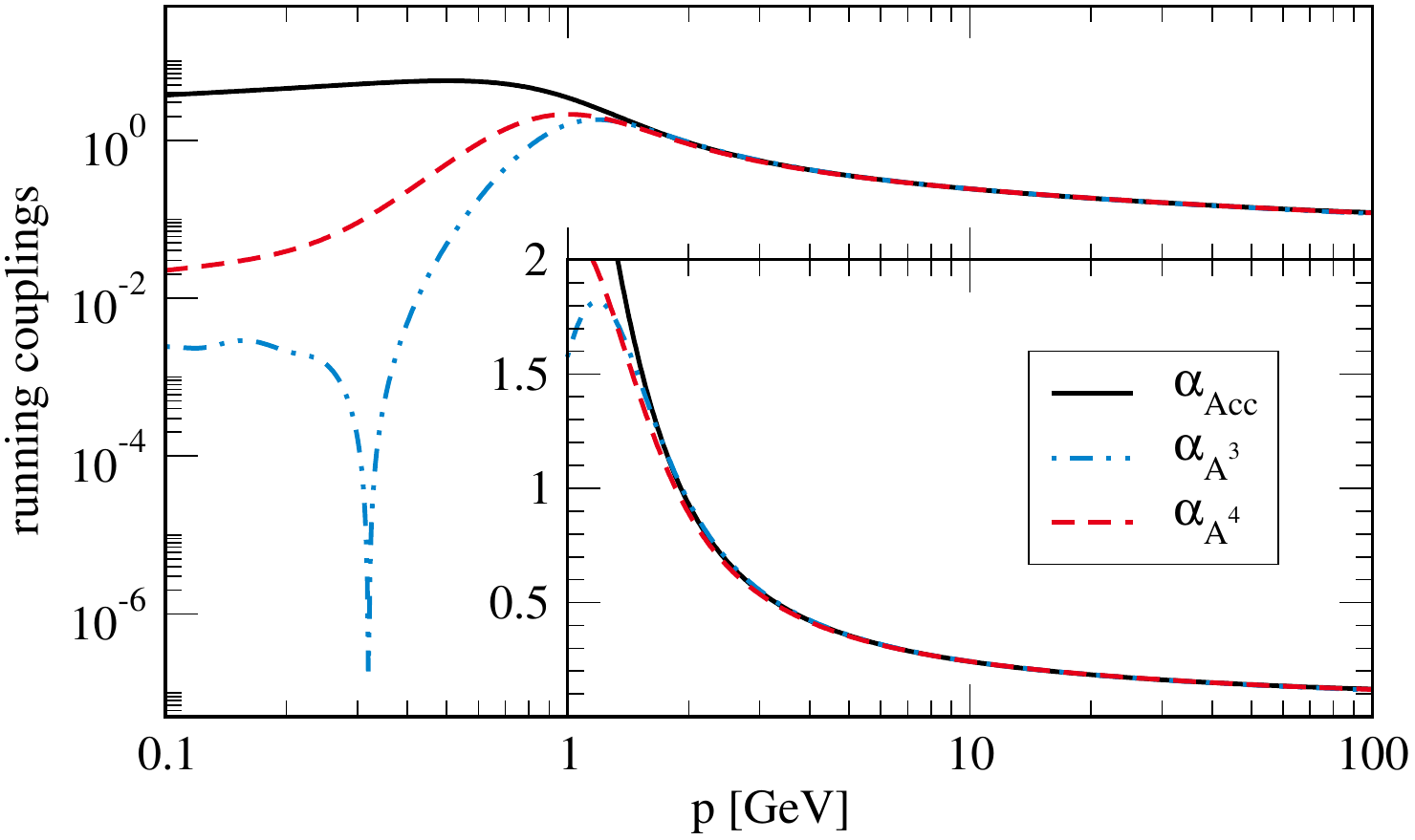}
        \caption{ Left: Gluon propagator in comparison to the lattice
          results from \cite{Sternbeck:2006cg}.  Right: Effective
          running couplings defined in \eq{eq:runcoup} as obtained
          from different Yang-Mills vertices as function of the
          momentum.
          }
     \label{fig:main_result_II}
\end{figure*}

\subsection{Origin of irregularities}

As discussed in the previous section, self-consistency in terms of the
Slavnov-Taylor identities entails a correspondence between the
dynamical generation of a gluon mass gap and the presence of
irregularities. But the STIs do not provide a mechanism for the
creation of irregularities, the gluon mass gap, and in turn
confinement.

In the scaling solution, \eq{eq:scaling_sol}, the irregularities arise
naturally from the non-trivial scaling. Hence they are tightly linked
to the original Kugo-Ojima confinement scenario \cite{Kugo:1979gm}, that requires the 
non-trivial scaling. Note, however, that this simply links different
signatures of confinement but does not reveal the mechanism at work.

For the decoupling solution \eq{eq:decoupling_sol}, we want to
discuss two possible scenarios. In the first scenario, the irregularities
are generated in the far infrared. A second possibility is that they are 
triggered via a condensate and/or a resonance, providing a direct connection 
of confinement and spontaneous symmetry breaking.

In the first scenario it is sufficient to focus on ghost loops as
possible sources of such irregularities, since the gluonic diagrams
decouple from the infrared dynamics due to the gluon mass gap.  This
is a seemingly appealing scenario as it is the dynamical ghost that
distinguishes confining Yang-Mills theory from e.g.\ QED. However, in
the decoupling solution \eq{eq:decoupling_sol} both, the ghost-gluon
vertex as well as the ghost propagator, have infrared finite quantum
corrections: no ghost-loops contribute to their equation and
(infrared) constant dressing functions can be assumed for both. As a
consequence the ghost loop contributions to correlation functions have
the same infrared structure as perturbative ghost-loop
contributions. However, none of these perturbative ghost loops yields
the necessary irregularities, see \App{app:ghosttriangle} for an
explicit calculation.

In the second scenario, the generation of irregularities can be based
on the dynamical generation of a non-vanishing transverse background,
$F_{\mu\nu}^a F_{\mu\nu}^a \neq 0\,$, in the infrared. This gluon
condensate is the Savvidi vacuum \cite{Savvidy:1977as}, and its
generation in the present approach has been discussed in
\cite{Eichhorn:2010zc} with $F_{\mu\nu}^a F_{\mu\nu}^a \approx
\SI{1}{\GeV}^4\,$. Then, a vertex expansion about this non-trivial
IR-solution of the equation of motion introduces an IR-splitting of
transverse and longitudinal vertices due to the transversality of the
background field. This IR-splitting automatically implies
irregularities as discussed in \Sec{sec:gluonmassirregularities}, and 
is sufficient for creating a physical mass gap in the gluon. This
scenario provides a direct relation of confinement and spontaneous
symmetry breaking. Therefore it is possibly connected to the presence
of resonances that are triggered in the longitudinal sector of the
theory, where they do not spoil the gapping of the completely
transverse sector. A purely longitudinal massless mode, as a source
for irregularities in the gluonic vertices, has been worked out in
\cite{Aguilar:2011ux,Aguilar:2011xe}, for a short summary see
\cite{Figueiredo:2016cvf}.  As a consequence, an irregularity appears
in the purely longitudinal three-gluon vertex in a way that preserves
the corresponding Slavnov-Taylor identity. The creation of a purely
transverse background and the presence of longitudinal massless mode
would then be two sides of the same coin. Furthermore, the
longitudinal resonance has to occur at about the same scale as the
gluon condensate, in order to trigger the correct gluon mass gap. A
more detailed discussion and computation of this scenario cannot be
assessed in the purely transverse system and is therefore deferred to
future work.
\begin{figure*}
	\centering
	\includegraphics[width=0.488\textwidth]{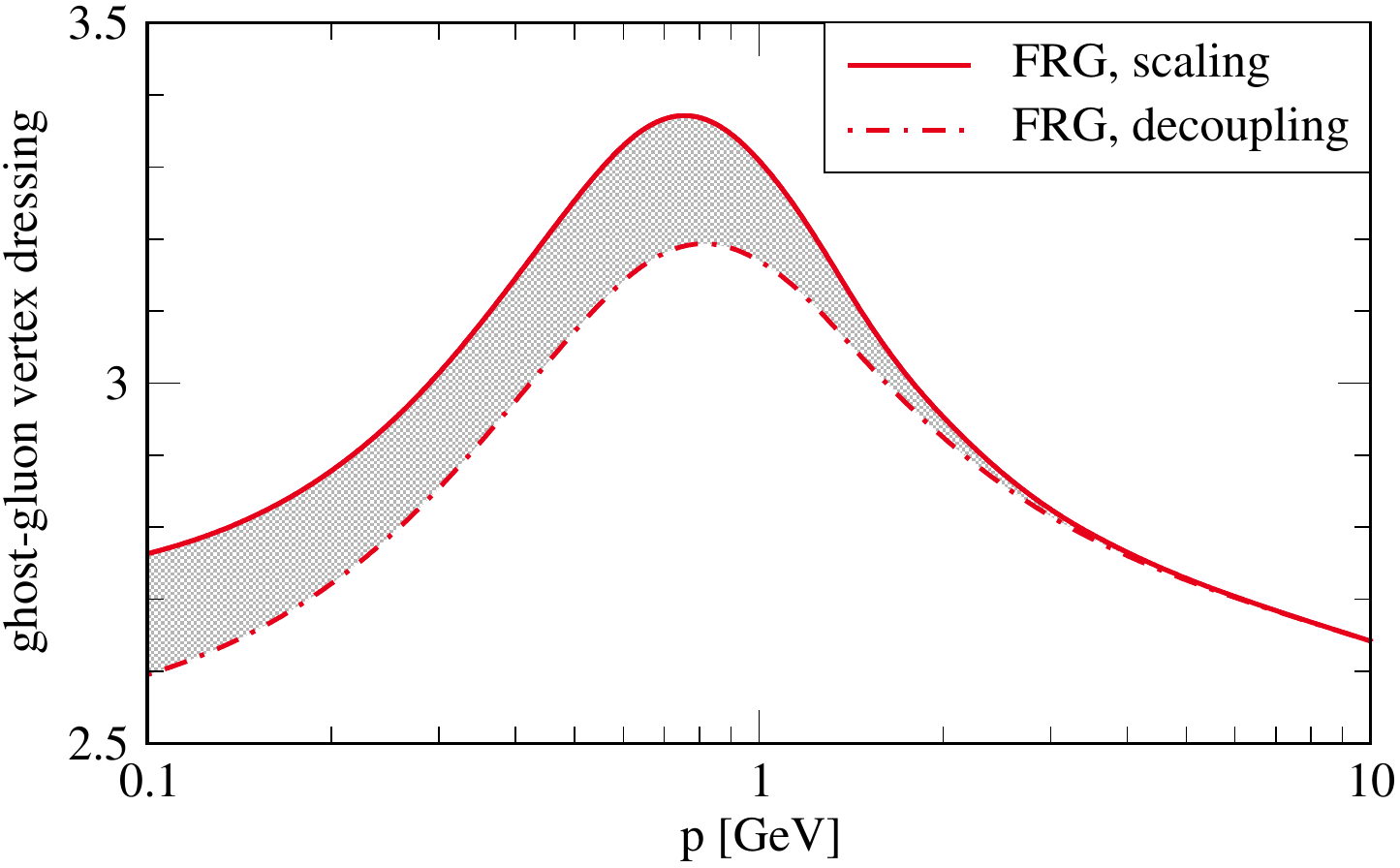}
\hfill
	\includegraphics[width=0.48\textwidth]{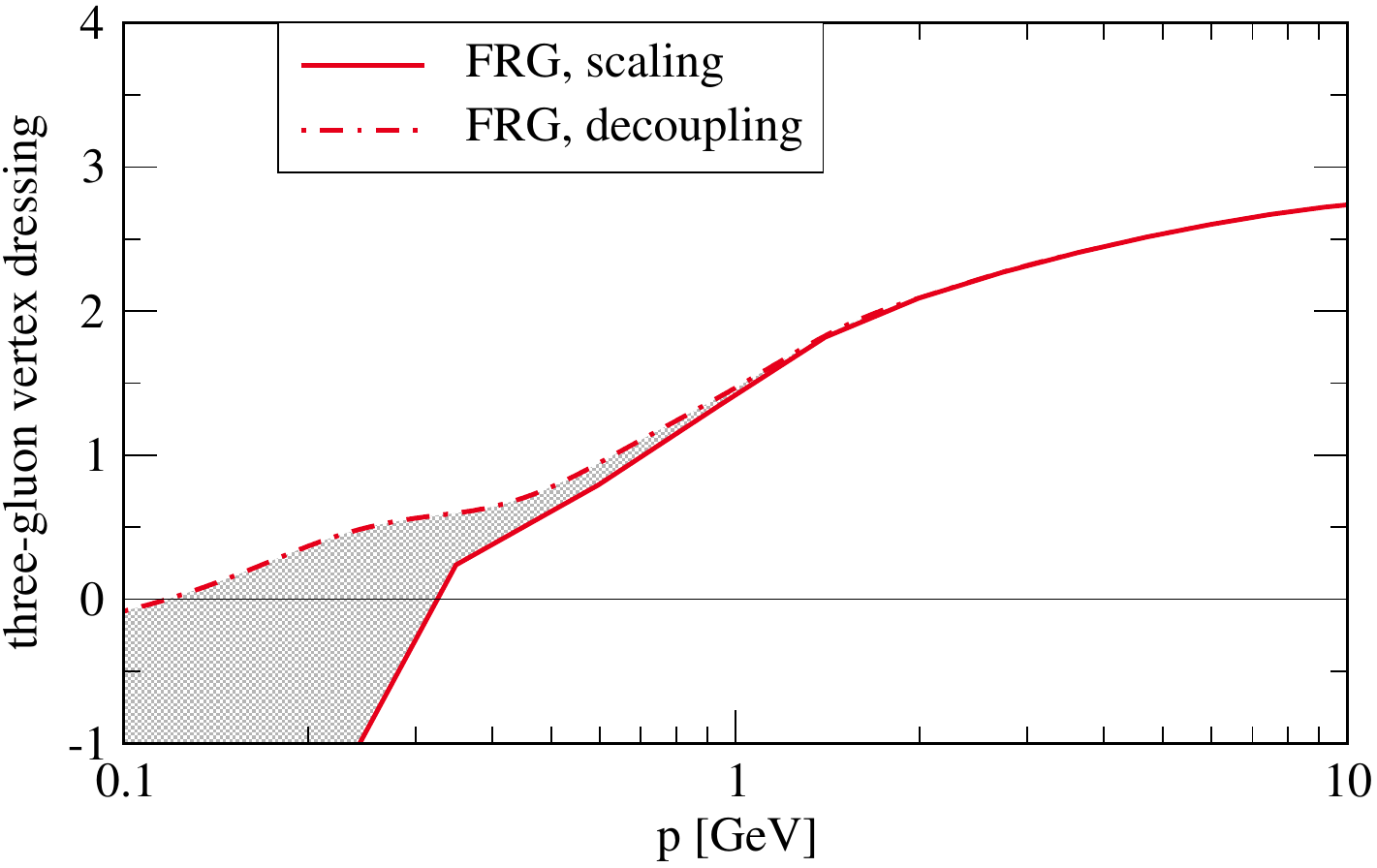}
	\caption{ Ghost-gluon vertex (left) and three-gluon
	  (right) vertex dressing functions $Z_{A \bar
	    cc,\bot }\left(p,\,p,\,-\frac{1}{2}\right)$ and
	  $Z_{A^3,\bot }\left(p,\,p,\,-\frac{1}{2}\right)$ in the
	  symmetric point configuration.  More momentum
	  configurations and comparisons to Dyson-Schwinger
	  and lattice results can be found in 
	  Figs.~\ref{fig:fourGluonTadpoleDressing}-\ref{fig:ThreePointVertices}.  
	  In contrast to \Fig{fig:main_result}, the decoupling
	  dressings are normalised to the scaling solution
	  in the UV. }
	\label{fig:main_result_III}
\end{figure*}

\subsection{The purely transverse system}
\label{sec:transverse}

In this work we restrict ourselves to a solution of the purely
transverse system \eq{eq:closedFun}, which is closed. The only
relevant UV parameters in this system are the strong coupling and the
transverse gluon mass parameter. In the UV the transverse mass
parameter agrees with the longitudinal one. The latter is fixed by the
mSTI for the longitudinal gluon propagator. Hence, the only
information needed from the longitudinal system is the initial value
for the transverse gluon mass parameter \eq{eq:mSTI_mass}. Note also
that there is at least one value for the initial gluon mass parameter
that yields a valid confining solution. In the following we vary the
gluon mass parameter and discuss the properties of the ensuing
solutions. We find a confining branch with both scaling and decoupling
solutions. In addition, we observe a transition to the deconfined
Higgs-type branch. No Coloumb branch is found. The unique scaling
solution satisfies the original Kugo-Ojima confinement criterion with
$Z_C(p=0)=0\,$. We emphasise that the existence of the scaling
solution is dynamically generated in a highly non-trivial way. The
details are discussed in \Sec{sec:gluonmassgap}.

\section{Numerical results}
\label{sec:mainresult}

We calculate Yang-Mills correlation functions by integrating the
self-consistent system of flow equations obtained from functional
derivatives of \eq{eq:flow}, see \Fig{fig:diagrams} 
for diagrammatic representations. Technical details on the
numerical procedure are given in \App{app:technicalDetails}.  We use
constant dressing functions as initial values for the $1$PI
correlators at the ultraviolet initial scale $\Lambda\,$. Consequently,
the initial action $\Gamma_\Lambda$ is given by the bare action of QCD
and the Slavnov-Taylor identities enforce relations between these
constant initial correlation functions.  As is well-known, and also
discussed in \Sec{sec:mSTIandVert_sub}, the Landau gauge STIs leave
only three of the renormalisation constants independent, namely the
value of the strong running coupling and two trivial renormalisations
of the fields that drop out of any observable.  To eliminate cutoff
effects, we choose the constant initial values for the vertex
dressings such that the momentum-dependent running couplings,
\eq{eq:runcoup} are degenerate at perturbative momentum scales $p$
with $\Lambda_\text{\tiny{QCD}}\ll p\ll \Lambda\,$ i.e. the
  STIs \eq{eq:RGrel} are only fulfilled on scales considerably below
  the UV cutoff scale. The modification of the Slavnov-Taylor
identity, caused by the regulator term, requires a non-physical gluonic mass 
term $m_\Lambda^2$ at the cutoff $\Lambda$. The initial value for 
the inverse gluon propagator is therefore taken as
\begin{align}
  [\Gamma^{(2)}_{AA,\Lambda}]^{ab}_{\mu\nu}(p) &=
  \left(Z_{A,\Lambda}\, p^2+ m_\Lambda^2\right)\,\delta^{ab}\,
  \Pi^{\bot}_{\mu\nu}(p)\,.
\end{align}
The non-physical contribution $m_k^2$ to the gluon propagator vanishes
only as the renormalisation group scale, $k\,$, is lowered to zero,
where the mSTIs reduce to the usual Slavnov-Taylor identities.  The
initial value $m_\Lambda^2$ can be uniquely fixed by demanding that
the resulting propagators and vertices are of the scaling
type. Consequently, the only parameter in this calculation is the
value of the strong running coupling at the renormalisation scale, as
initially stated.  We also produce decoupling solutions by varying the
gluon mass parameter towards slightly larger values.  Our reasoning
for their validity as confining solutions is presented in
\Sec{sec:gluonmassgap}.
\begin{figure*}
	\centering
	\includegraphics[width=0.497\textwidth]{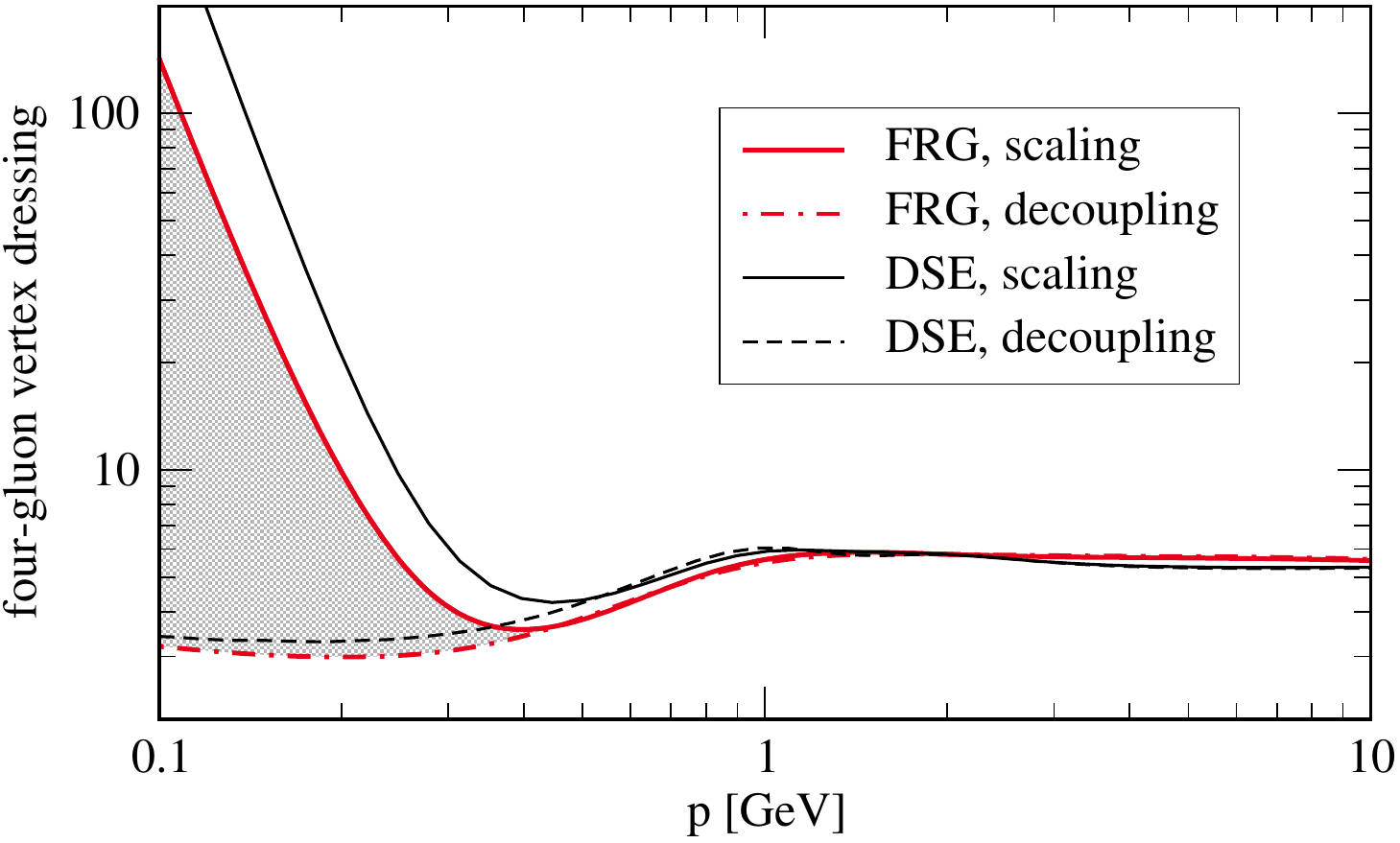}
	\hfill
	\includegraphics[width=0.48\textwidth]{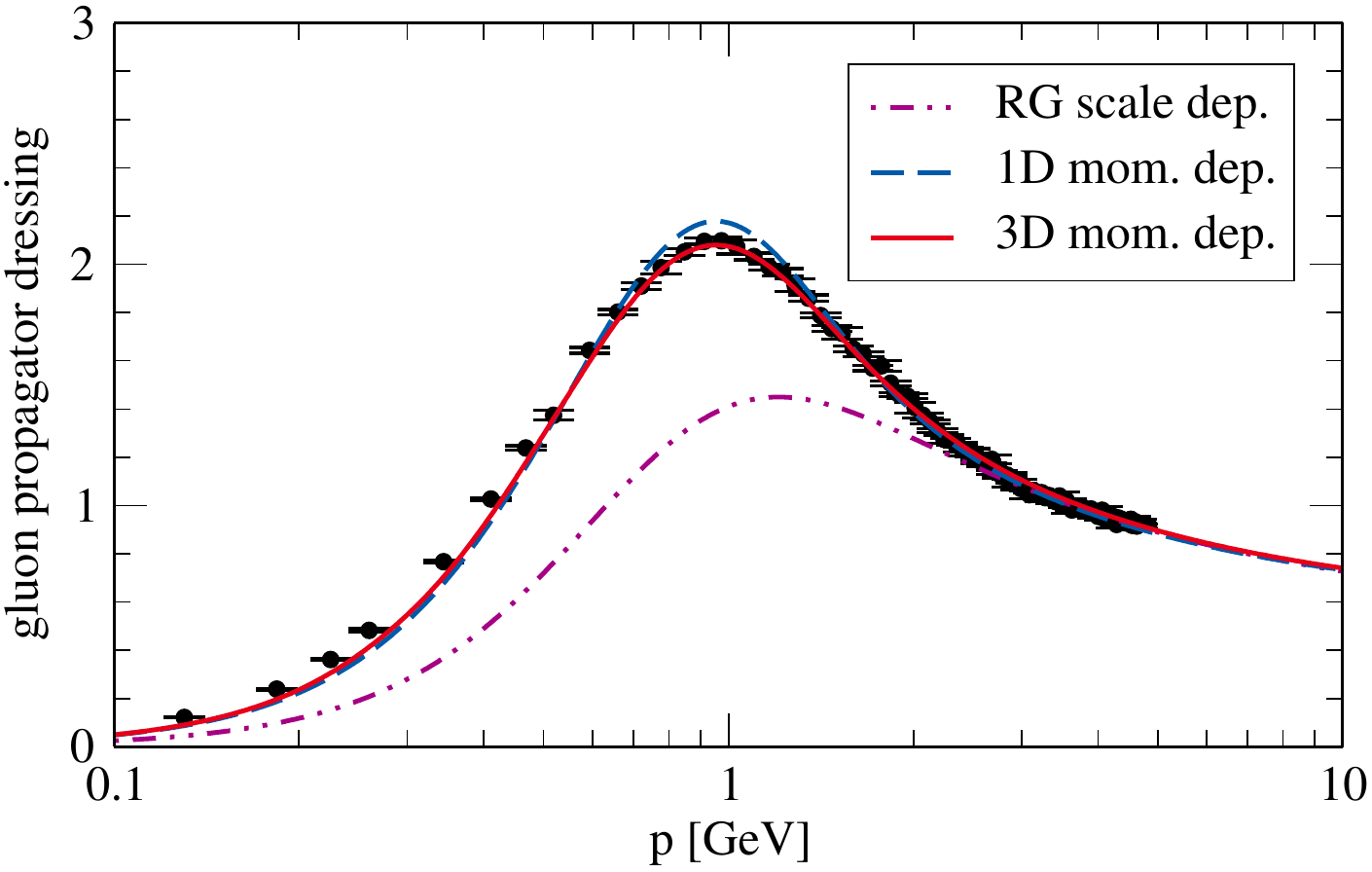}
        \caption{ Left: 
	  Four-gluon vertex dressing function as defined in
          \eq{eq:fourgluon} at the symmetric point in comparison to
          Dyson-Schwinger computations~\cite{Cyrol:2014kca}.  We
          normalised all curves to match the scaling result at
          $p=\SI{2}{\GeV}$. Right: Gluon propagator dressings obtained with
          different momentum approximations, see
          \Sec{sec:truncationcheck} for details.
         }
	\label{fig:main_result_IIII}
\end{figure*}

\subsection{Correlation functions and running couplings}
\label{sec:corcoup}

We show our results for the Yang-Mills correlation functions as well
as the momentum-dependent transverse running couplings in
Figs.~\ref{fig:main_result}-\ref{fig:main_result_IIII}, see also
Figs.~\ref{fig:fourGluonTadpoleDressing}-\ref{fig:ThreePointVertices}
in the appendices for a comparison of the vertices to recent lattice
and DSE results.  A discussion of truncation effects is deferred to
\Sec{sec:truncationcheck}.  In order to be able to compare to results
from lattice simulations, we set the scale and normalise the dressings
as described in \App{app:rescaling}.  At all momenta, where the
difference between the scaling (solid line) and decoupling (band
bounded by dashed-dot line) solutions is negligible, our results for
the correlations functions agree very well with the corresponding
lattice results.  In the case of the scaling solution we find the
consistent scaling exponents
\begin{align}\nonumber 
 \kappa_\text{ghost}&=0.579\pm0.005\,,\\[2ex]
 \kappa_\text{gluon}&=0.573\pm0.002\,,
\end{align}
where the uncertainties stem from a least square fit with the ansatz
\begin{align}\nonumber 
 Z_c(p)&\varpropto (p^2)^{ \kappa_\text{ghost}}\,,\\[2ex]
 Z_A(p)&\varpropto (p^2)^{-2\, \kappa_\text{gluon}}\,.
\end{align}

As discussed in \Sec{sec:transverse}, the scaling solution is a
self-consistent solution of the purely transverse system in the used
approach, and has no systematic error related to the lack of solving
the longitudinal system. In turn, the presented decoupling solutions
suffer from the missing solution of the longitudinal system, leading
to a small additional systematic error.  This argument already
suggests that it is the presented scaling solution that should agree
best with the lattice results in the regime $p\gtrsim \SI{1}{\GeV}\,$,
where the solutions show no sensitivity to the Gribov problem. This is
confirmed by the results, see in particular \Fig{fig:main_result}.

In the infrared regime, $p\lesssim \SI{1}{\GeV}\,$, the different
solutions approach their infrared asymptotics. In
\Fig{fig:main_result} and \Fig{fig:main_result_II} we compare the FRG
solutions with the lattice data from \cite{Sternbeck:2006cg}. In
agreement with other lattice results
\cite{Cucchieri:2007rg,Cucchieri:2008fc,Maas:2009ph} in four
dimensions, these propagators show a decoupling behaviour, for a
review see \cite{Maas:2011se}. Taking the IR behaviour of all
correlators into account, cf.\ also \Fig{fig:ThreePointVertices}, the
lattice solution \cite{Sternbeck:2006cg} is very close to the
decoupling solution (dot-dashed line) that is furthest from the
scaling solution (solid line). Note however, that the systematic
errors of both approaches, FRG computations and lattice simulations
increase towards the IR. While the FRG computations lack apparent
convergence in this regime, the lattice data are affected by the
non-perturbative gauge fixing procedure, i.e.\ the choice of Gribov
copies \cite{Maas:2009se,Sternbeck:2012mf,Maas:2015nva} and
discretisation artefacts \cite{Duarte:2016iko}.  Consequently,
comparing the FRG IR band to the lattice propagators has to be taken
with a grain of salt.  In the case of the vertices, we compare also to
results obtained within the Dyson-Schwinger equation approach
\cite{Huber:2012kd,Blum:2014gna,Cyrol:2014kca}, see
\Fig{fig:comparison} and \ref{fig:ThreePointVertices}. A comparison of
the different running vertex couplings is given in
\Sec{sec:comparison}.

We find that it is crucial to ensure the degeneracy in the different
running vertex couplings at perturbative momentum scales in order to
achieve quantitative accuracy, see also \Sec{sec:truncationcheck}.
The transverse effective running couplings, as defined in
\eq{eq:runcoup}, are shown in the right panel of
\Fig{fig:main_result_II}.  To be able to cover a larger range of
momenta with manageable numerical effort, the shown running couplings
have been obtained within an approximation that takes only one
momentum variable into account in the vertices, see
\Sec{sec:truncationcheck}.  At large perturbative momentum scales, we
find them to be perfectly degenerate, as is demanded by the
Slavnov-Taylor identities.  The degeneracy of the running couplings is
lifted at a scale of roughly $\SI{2}{\GeV}\,$, which coincides with the
gapping scale of the gluon.  Furthermore, the three-gluon vertex shows
a zero crossing at scales of \SIrange{0.1}{0.33}{GeV}, which is the
reason for the spike in the corresponding running coupling.  This zero
crossing, which is caused by the infrared-dominant ghost-loop, is
well-known in the literature
\cite{Aguilar:2013vaa,Pelaez:2013cpa,Blum:2014gna,Eichmann:2014xya}.
Even though we are looking at the scaling solution, we find that the
running couplings defined from the purely gluonic vertices are still
strongly suppressed in the infrared. In particular the three-gluon
vertex running coupling becomes more strongly suppressed than the
four-gluon vertex running coupling.  However, as demanded by scaling,
they seem to settle at tiny but finite fixed point values, which has
also been seen in Dyson-Schwinger studies
\cite{Eichmann:2014xya,Kellermann:2008iw,Cyrol:2014kca}.

\subsection{Quality of the approximation}
\label{sec:truncationcheck}

In \Fig{fig:main_result_IIII} (right panel), we show the scaling
solution for the propagators in different truncations.  In all cases,
the full momentum dependence of the propagators is taken into account
whereas different approximations are used for the vertices. Including
only RG-scale-dependent constant vertex dressing functions is the
minimal approximation that can produce a scaling solution with a
physical gluon mass gap.  The dot-dashed (magenta) line in
\Fig{fig:main_result_IIII} (right panel) corresponds to an
approximation with constant vertex dressing functions evaluated at the
symmetric configuration with momentum $\mathcal{O}(\SI{250}{\MeV})\,$.
Hence the vertices are only RG-scale-dependent vertices. For the
dashed blue results the dressing functions for the transversally
projected classical tensor structures have been approximated with a
single momentum variable $\bar{p}^2 \equiv \tfrac{1}{n}\sum_{i=1}^n
p_i^2\,$. Reducing the momentum dependence to a single variable
requires the definition of a momentum configuration to evaluate the
flow.  Here, we use the symmetric point configuration, defined by $p_i
\cdot p_i = p^2$ and $p_i\cdot p_j = -1/(n-1)$ for $i\neq j\,$, where
$n=3\, (4)$ for the three(four)-gluon vertex. Finally, the solid red
line corresponds to our best truncation. As described in
\Sec{sec:expsch}, it takes into account the full momentum dependence
of the classical tensors structures of the three-point functions as
well as the four-gluon vertex in a symmetric point
approximation. Additionally, all (three-dimensional) momentum
configurations of the four-gluon vertex that are needed in the tadpole
diagram of the gluon propagator equation have been calculated and
coupled back in this diagram.  The reliability of our approximation
can be assessed by comparing the two simpler truncations to the result
obtained in our best truncation scheme. We observe that our results
apparently converge towards the lattice result, as we improve the
momentum approximation for the vertices.

The effects of non-classical tensor structures and vertices are beyond
the scope of the current work and have to be checked in future
investigations, see however \cite{Eichmann:2014xya} for an
investigation of non-classical tensor structures of the three-gluon
vertex. Within the present work, the already very good agreement with
lattice results suggests, that their influence on the propagators is
small.

The final gluon propagator is sensitive to the correct renormalisation
of the vertices. For example, a one percent change of the three-gluon
vertex dressing at an UV scale of $\SI{20}{\GeV}$ magnifies by up to a
factor 10 in the final gluon propagator. Therefore, small errors in
the perturbative running of the vertices propagate, via
renormalisation, into the two-point functions. We expect a five
percent uncertainty in our results due to this.

Despite these uncertainties, we interpret the behaviour in
\Fig{fig:main_result_IIII} (right panel) as an indication for apparent
convergence.

\subsection{Comparison to other results}
\label{sec:comparison}
\begin{figure}
	\includegraphics[width=0.48\textwidth]{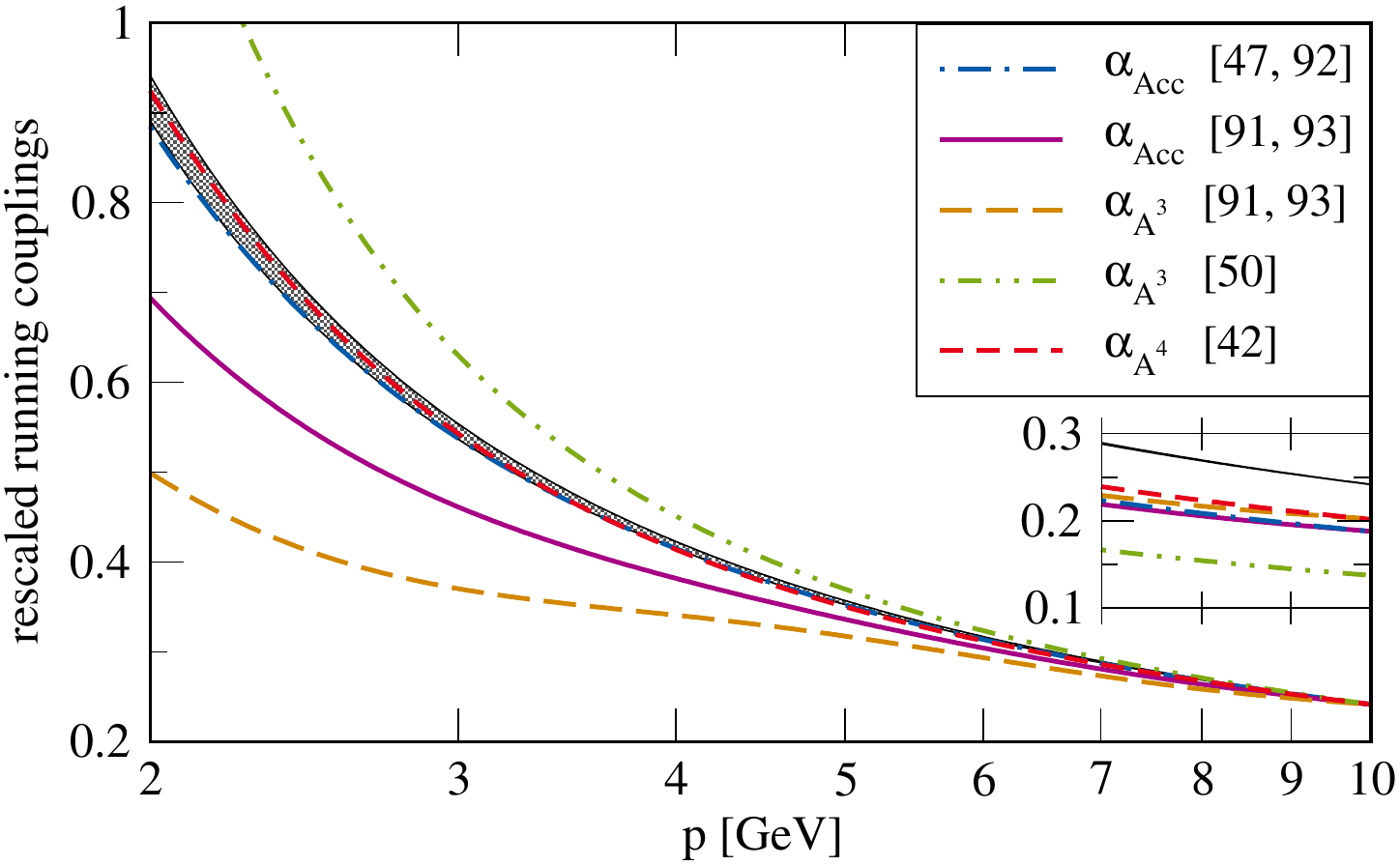}
	\caption{Running couplings \eq{eq:runcoup} in comparison with DSE running.  
	  The grey band gives the spread of vertex
          couplings from the FRG in the present work.  The DSE results
          are shown rescaled to fit our ghost-gluon vertex running
          coupling at $\SI{10}{\GeV}$ to facilitate the comparison.
          The inlay shows the unscaled couplings.  Note that the FRG
          running couplings naturally lie on top of each other and are
          not depicted rescaled.}
	\label{fig:comparison}
\end{figure}

In \Fig{fig:ThreePointVertices}, numerical results for the ghost-gluon and
three-gluon vertices are shown in comparison to other functional
methods as well as lattice results. In summary, the results from
various functional approaches and the lattice agree to a good degree.
But these correlation functions are not renormalisation group
invariant, and a fully meaningful comparison can only be made with
RG invariant quantities.  Therefore, we compare our results for the
RG invariant running couplings with the respective results from DSE
computations. To be more precise, it is actually the $\beta$ functions
of the different vertices that are tied together by two-loop
universality in the sense that they should agree in the regime where
three-loop effects are negligible. Since constant factors drop out of
the $\beta$ functions, we have normalised the DSE running couplings to
the FRG result at large momentum scales in \Fig{fig:comparison}.
For the sake of visibility, we only have provided a band for
the spread of the FRG couplings as obtained from different vertices.
The shown DSE running couplings are based on a series of works
\cite{Huber:2012kd,Blum:2014gna,
  Cyrol:2014kca,Williams:2014iea,Williams:2015cvx}, where the explicitly
shown results are taken from
\cite{Blum:2014gna,Cyrol:2014kca,Huber:unpublished,Williams:unpublished}.
Additionally, we provide the raw DSE running couplings that have not
been rescaled by a constant factor in the inlay.

\subsection{Mass gap, mSTIs and types of solutions}
\label{sec:gluonmassgap}
\begin{figure*}
	\centering
	\includegraphics[width=0.481\textwidth]{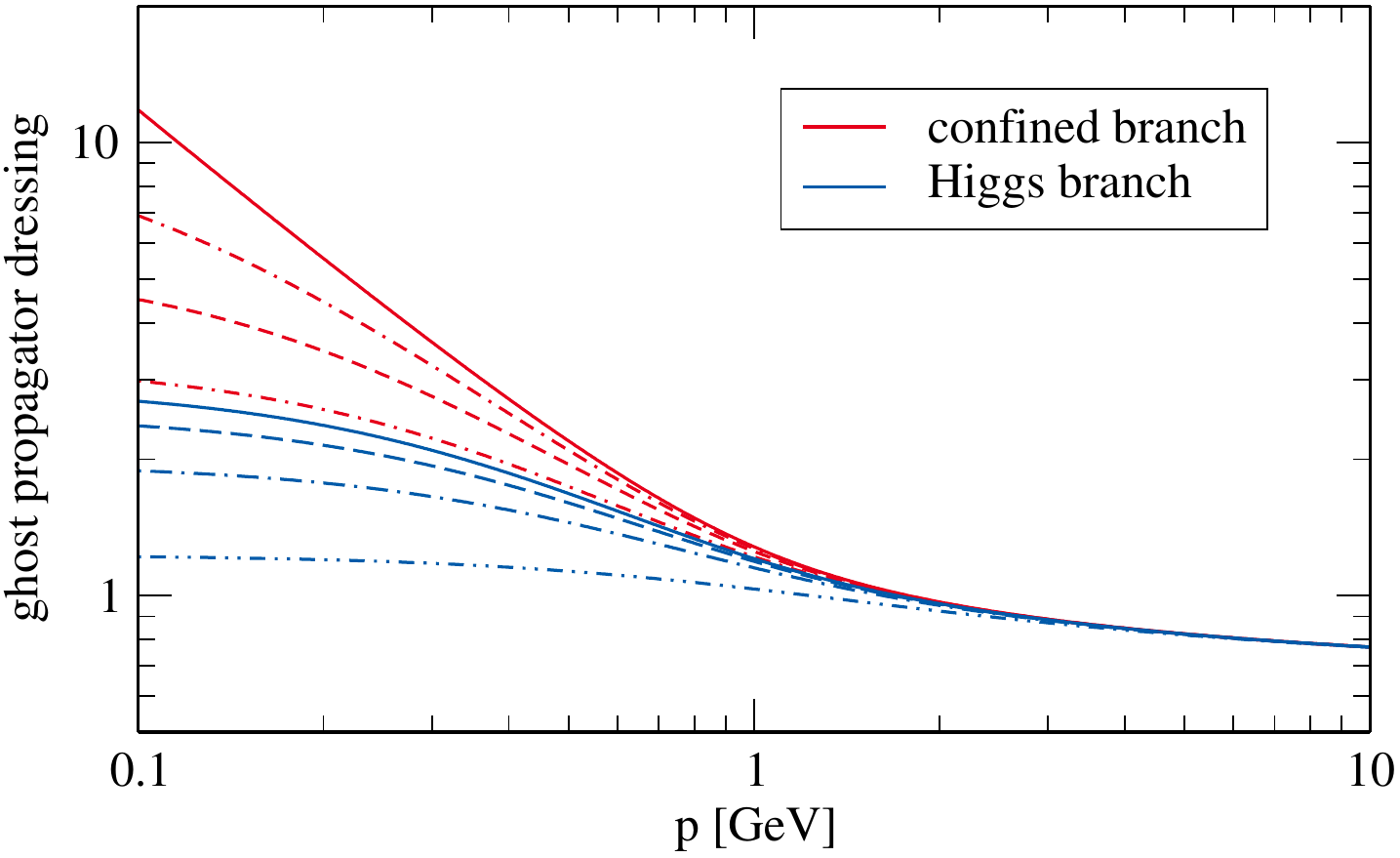}
\hfill
	\includegraphics[width=0.48\textwidth]{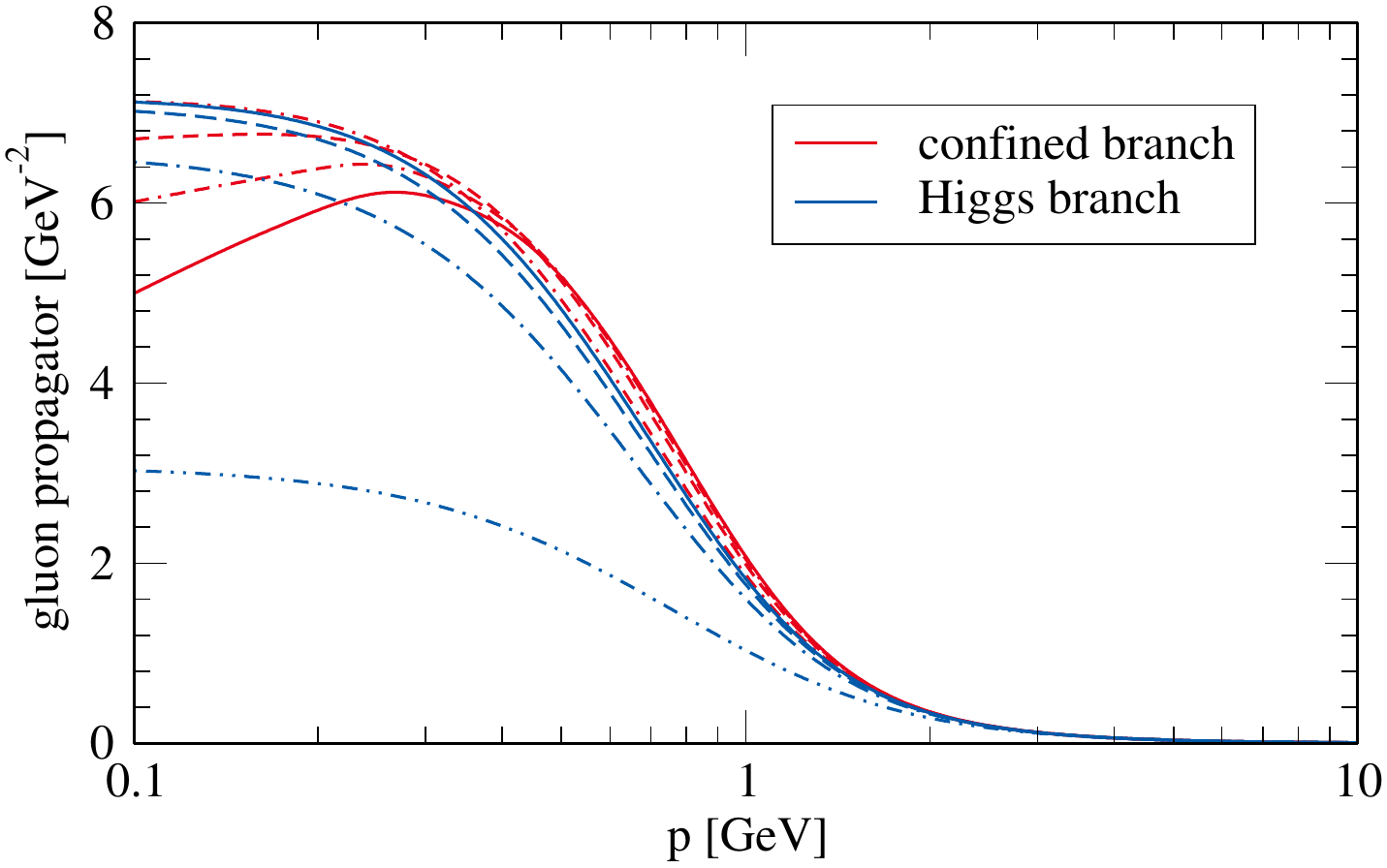}
        \caption{Ghost dressing functions $1/Z_c$ (left) and
          gluon propagators (right) for different values of the
          ultraviolet gluon mass parameter. Blue results correspond to
          the Higgs-type branch and red results to the confined branch. The
          solutions in all branches have been normalised to the
          scaling solution in the UV. }
	\label{fig:confinementVSHiggs}
\end{figure*}

As discussed in \Sec{sec:mSTIandVert_sub}, the introduction of the
regulator in the FRG leads to a modification of the Slavnov-Taylor
identities. In turn the inverse gluon propagator obtains a
contribution proportional to $\Delta\Gamma^{(2)}_{AA}\varpropto
k^2\alpha(k)$ for all $k>0\,$. Disentangling the physical mass gap
contribution from this mSTI contribution to the gluon mass parameter
is intricate, both conceptually and numerically. The resulting
numerical challenge is illustrated in the appendix in
\Fig{fig:flowQuantities}, where we show the $k$-running of the gluon
mass parameter. This is the analogue of the problem of quadratic
divergences in Dyson-Schwinger equations with a hard momentum cutoff,
see e.g.\ \cite{Huber:2014tva}.  However, there has to exist at least
one choice for the gluon mass parameter $m_\Lambda^2$ that yields a
valid confining solution, see \Sec{sec:mSTIandVert_sub}.  To resolve
the issue of finding this value, we first recall that a fully regular
solution has no confinement and necessarily shows a Higgs- or
Coulomb-type behaviour. Although we do not expect these branches to be
consistent solutions, we can trigger them by an appropriate choice of
the gluon mass parameter in the UV. The confinement branch then lies
between the Coulomb and the Higgs branch. We need, however, a
criterion for distinguishing between the confinement and the
Higgs-type branch.

\begin{figure*}
	\centering
	\includegraphics[width=0.485\textwidth]{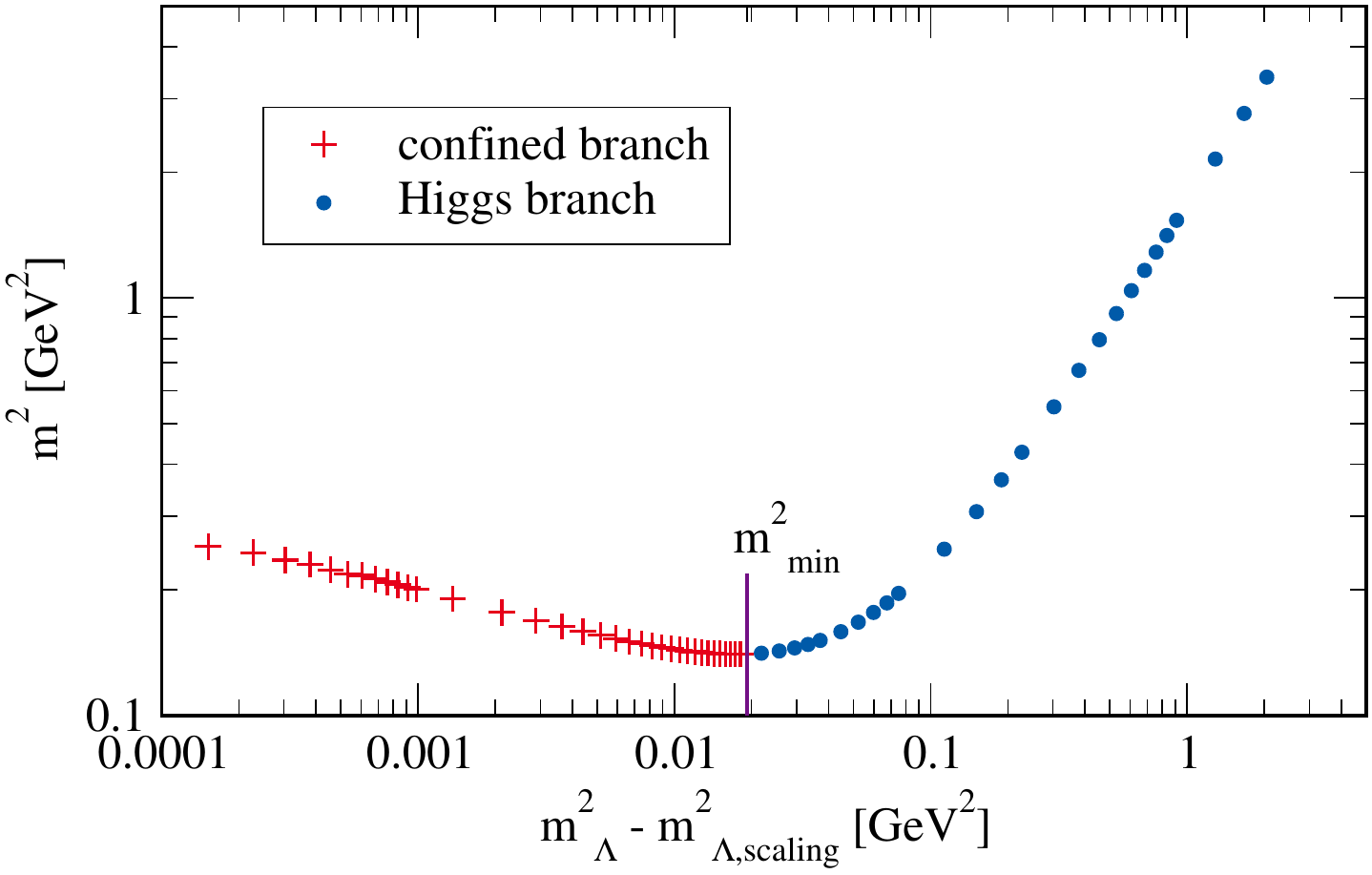}
        \hfill
	\includegraphics[width=0.48\textwidth]{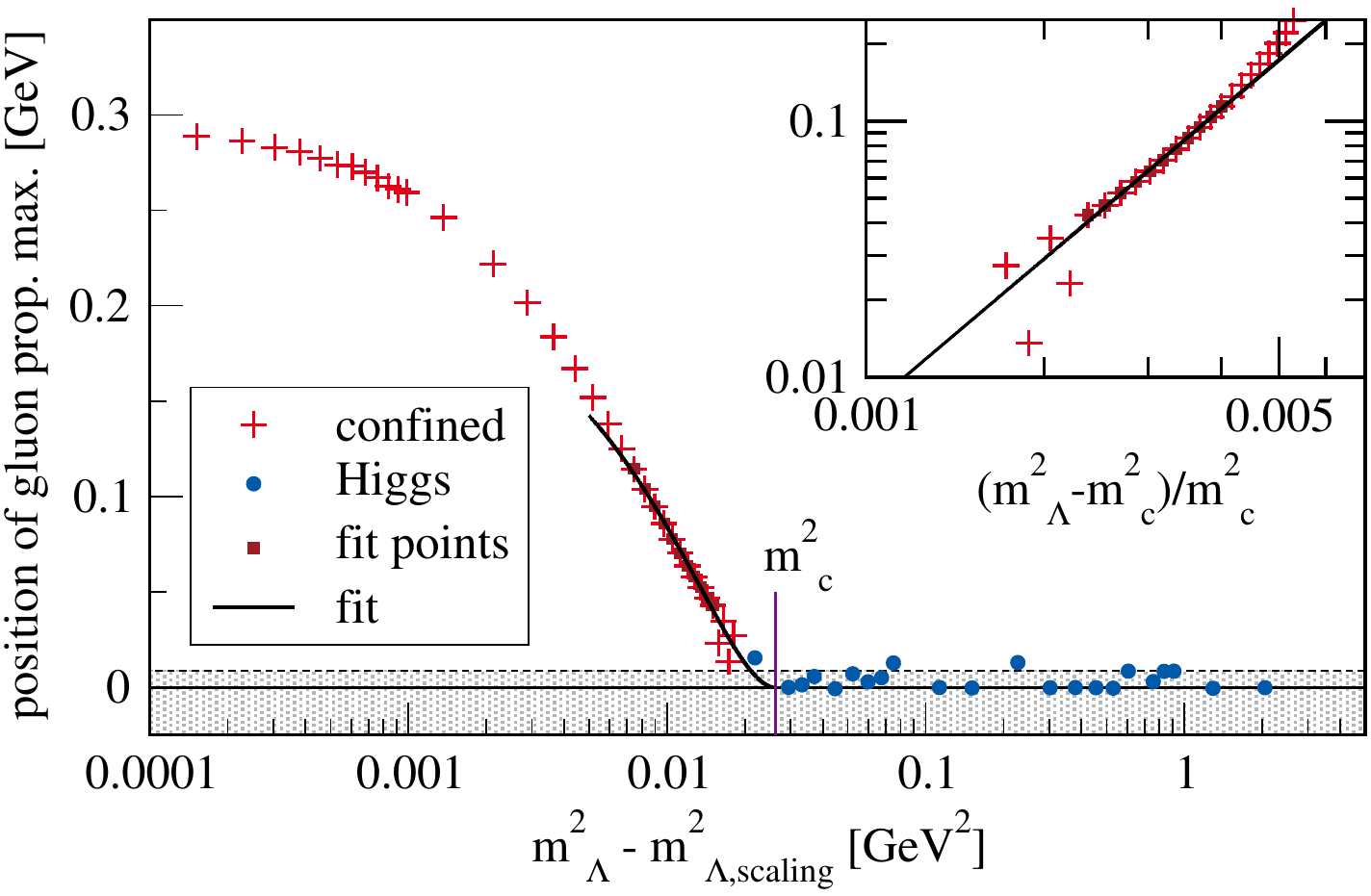}
        \caption{ Left: Gluon mass gap as a function of the gluon mass
          parameter $m_\Lambda^2-m_\text{$\Lambda$, scaling}^2\,$,
          where $m_\text{$\Lambda$, scaling}^2$ denotes the gluon mass
          parameter that yields the scaling solution. Right: Momentum
          value at which the gluon propagator assumes its maximum, as
          a function of the gluon mass parameter
          $m_\Lambda^2-m_\text{$\Lambda$, scaling}^2\,$.  The inlay
          exposes the power law behaviour of the gluon propagator
          maximum in the vicinity of the transition region, see
          \eq{eq:phaseTransitionFit}. Both plots were obtained from
          our numerically less-demanding 1D approximation. We have
          repeated this analysis in the transition regime from
          Higgs-type to confinement branch also with the best
          approximation and find the same behaviour.  The shaded area
          marks momentum scales that are not numerically resolved in
          the present work.  The points in this region rely on a
          generic extrapolation.  }
	\label{fig:mass_parameter}
\end{figure*}

To investigate the possible solutions in a controlled way, we start
deep in the Higgs-type branch: an asymptotically large initial gluon
mass parameter $m_\Lambda^2$ triggers an explicit mass term of the
gluon at $k=0\,$. If we could trigger this consistently in the present
$SU(3)$ theory, it would constitute a Higgs solution. Note that in the
current approximation it cannot be distinguished from massive
Yang-Mills theory, which has e.g.\ been considered in
\cite{Tissier:2010ts,Pelaez:2014mxa}. Starting from this Higgs-type
branch, we can then explore the limit of smaller initial mass
parameters.  This finally leads us to the scaling solution, which
forms the boundary towards an unphysical region characterised by
Landau-pole-like singularities. It is left to distinguish between the
remaining confining and Higgs-type solutions, shown in
\Fig{fig:confinementVSHiggs}, without any information from the
longitudinal set of equations. For that purpose we use two criteria:

In the left panel of \Fig{fig:mass_parameter}, we show the mass gap of
the gluon, $m^{2}=\Gamma^{(2)}_{AA,k=0}(p=0)\,$, as a function of the
chosen initial value for the gluon mass parameter $m_{\Lambda}^2$
subtracted by the corresponding value for the scaling solution
$m_{\Lambda,\rm scaling}^2\,$.  The latter solution corresponds to zero
on the x-axis in \Fig{fig:mass_parameter}. As mentioned before, going
beyond the scaling solution, $m_{\Lambda}^2<m_{\Lambda,\rm
  scaling}^2\,$, leads to singularities.  We interpret their presence as
a signal for the invalidity of the Coulomb branch as a possibly
realisation of non-Abelian Yang-Mills theory.  The decisive feature of
the left panel of \Fig{fig:mass_parameter} is the presence of a
minimum at $m_{\rm min}^2\,$.  If there were no dynamical mass gap
generation, $m^2$ would have to go to zero as we lower
$m_{\Lambda}^2\,$. In contrast to this, we find that the resulting
gluon mass gap is always larger than the value it takes at
$m_{\Lambda}^2=m_{\rm min}^2\,$.  In particular, this entails that all
solutions to the left of the minimal value, $m_{\Lambda}^2<m_{\rm  min}^2\,$,
are characterised by a large dynamical contribution to the
gluon mass gap, which we interpret as confinement.

As a second criterion for differentiating between confining and Higgs
solutions, we use the presence of a maximum at non-vanishing momenta
in the gluon propagator, which signals positivity violation
\cite{Alkofer:2000wg}. In the right panel of \Fig{fig:mass_parameter},
we show the location of the maximum in the gluon propagator, again as
a function of the gluon mass parameter, $m_{\Lambda}^2-m_{\Lambda,\rm
  scaling}^2\,$. We clearly see a region of confining solutions that
show a back-bending of the gluon propagator at small momenta,
see \Fig{fig:confinementVSHiggs}.  The dashed line,
separating the shaded from the white region in the right panel of
\Fig{fig:mass_parameter}, indicates the smallest momentum value at
which the gluon propagator has been calculated.  With this restriction
in mind, the fit in the inlay demonstrates that the location of the
maximum of the propagator scales to zero as one approaches the
critical value $m_{\rm c}^2\,$. We fit with
\begin{align}
	\label{eq:phaseTransitionFit}
	p_\text{max}(m^2_\Lambda) \propto
        \left(\frac{m^2_\Lambda-m^2_{\rm c}}{m^2_{\rm
              c}}\right)^\alpha\,,
\end{align}
which yields the critical exponent
\begin{align}
	\label{eq:phaseTransitionFitResult}
  \alpha = 1.95 \pm 0.6\,, 
\end{align}
in the 1D approximation. Within the numerical accuracy, this boundary
value $m_{\rm c}^2$ is equivalent to the minimal value $m_{\rm min}^2$
of our first confinement criterion.  Hence, the value of the UV mass
parameter that results in the minimal gluon mass gap, is also the one
that shows minimal back-bending. Note that the lattice simulations
show a gluon propagator that is at least very close to this minimal
mass gap.

As discussed in detail in \Sec{sec:massgap} and appendices
\ref{app:irregularities} and \ref{app:ghosttriangle}, a gluon mass gap
necessitates irregularities.  The scaling solution by definition
contains these irregularities already in the propagators, cf.\
\eq{eq:scaling_sol}.  For the decoupling-type solutions, we excluded
infrared irregularities of diagrammatic origin, see
\App{app:ghosttriangle}.  Thus, for the decoupling-type solutions our
arguments for the validity of the solutions are weaker and remain to
be investigated in a solution including at least parts of the
longitudinal system, see the discussion in \Sec{sec:massgap}.
Additionally, it might be necessary to expand about the solution of
the equation of motion, see \cite{Eichhorn:2010zc}.  

We summarise the findings of the present section. In the right panel
of \Fig{fig:mass_parameter} we can distinguish a confining branch with
positivity violation and a Higgs-type branch with a massive gluon
propagator. A Coulomb-type solution, on the other hand, can never be
produced with the functional renormalisation group since any attempt
to do so leads to Landau-pole-like singularities. The non-existence of
the Coulomb branch is tightly linked to the non-monotonous dependence
of the mass gap on the initial gluon mass parameter, see left panel of
\Fig{fig:mass_parameter}. This behaviour is of a dynamical origin that is
also responsible for the existence of the scaling solution for the
smallest possible UV gluon mass parameter.

\subsection{Discussion}
\label{sec:discussion}

As has been discussed already in \Sec{sec:corcoup}, one non-trivial
feature of the different vertex couplings is their quantitative
equivalence for momenta down to $p\approx \SI{2}{\GeV}\,$, see
\Fig{fig:main_result_II} (right panel).  This property extends the
universal running of the couplings into the semi-perturbative regime.
On the other hand, the couplings violate universality in the
non-perturbative regime for $p\lesssim \SI{2}{\GeV}\,$.  The
universality down to the semi-perturbative regime is a very welcome
feature of Landau gauge QCD, as it reduces the size of the
non-perturbative regime and hence the potential systematic errors.  In
particular, one running coupling is sufficient to describe Landau
gauge Yang-Mills theory down to momentum scales of the order of the
gluon mass gap.  This suggests to use the propagators together with
the ghost-gluon vertex for simple semi-quantitative calculations.  The
above structure also explains and supports the semi-quantitative
nature of the results in low-order approximations.

This implies that self-consistent calculations of most or all vertices
have to reproduce this universality, in particular for momenta
$\SI{2}{\GeV}\lesssim p\lesssim \SI{10}{\GeV}\,$. When starting from
the value of the strong running coupling at perturbative momenta, we
find that a violation of the degeneracy of the running couplings,
\eq{eq:runcoup}, in this regime goes hand in hand with the loss of
even qualitative properties of the non-perturbative results in
self-consistent approximations.  This surprising sensitivity to even
small deviations of the couplings from their universal running extends
to the fully dynamical system with quarks, see e.g.\ 
\cite{Mitter:2014wpa,CMPS:VC}. Note in this context that the
quark-gluon coupling $\alpha_{A\bar q q}\,$, \eq{eq:quarkgluon},
agrees with the ghost-gluon coupling $\alpha_s$ defined in
\eq{eq:propcoupling}, and not the vertex coupling $\alpha_{A\bar c
  c}\,$, see \Sec{sec:STImSTI}. It can be shown in the full
QCD system, that deviations from universality on the percent level
have a qualitative impact on chiral symmetry breaking. The origin of
this is the sensitivity of chiral symmetry breaking to the correct
adjustment of physical scales, i.e.\ $\Lambda_{\text{\tiny QCD}}\,$,
in all subsystems.  These observations underline the relevance of the
present results for the quantitative grip on chiral symmetry
breaking. A full analysis will be presented in a forthcoming work,
\cite{CMPS:VC}.

We close this discussion with the remark that universality in the
semi-perturbative regime is tightly linked with the consistent
renormalisation of all primitively divergent correlation functions.
We find it crucial to demand the validity of the STIs \eq{eq:RGrel}
only on momentum scales considerably below the ultraviolet cutoff
$\Lambda\,$.  On the other hand, the relations \eq{eq:RGrel} are
violated close to the ultraviolet cutoff, due to the BPHZ-type
subtraction schemes. This constitutes no restriction to any practical
applications, since the cutoff can always be chosen large enough, such
that no violations effects can be found at momenta $p\ll \Lambda\,$.
One particular consequence of BPHZ-type subtraction schemes is then
that the calculated renormalisation constants necessarily have to
violate \eq{eq:RGrel}, since they contain contributions from momentum
regions close to the ultraviolet cutoff.

\section{Conclusion}

In this work we investigate correlation functions in Landau gauge
$SU(3)$ Yang-Mills theory.  This analysis is performed in a vertex
expansion scheme for the effective action within the functional
renormalisation group approach.  Besides the gluon and ghost
propagators, our approximation for the effective action includes the
self-consistent calculation of momentum-dependent dressings of the
transverse ghost-gluon, three-gluon and four-gluon vertices.  Starting
from the gauge fixed tree-level perturbative action of Yang-Mills
theory, we obtain results for the correlators that are in very good
agreement with corresponding lattice QCD simulations. Furthermore, the
comparison of different vertex truncations indicates the apparent
convergence of the expansion scheme.

Special emphasis is put on the analysis of the dynamical creation of
the gluon mass gap at non-perturbative momenta. Self-consistency in
terms of the Slavnov-Taylor identities directly links this property to
the requirement of IR irregularities in the correlation functions. The
source of these irregularities is easily traced back to the
IR-divergent ghost propagator for the scaling solution.  In the
decoupling-type solutions, the source of these irregularities is
harder to identify, where the creation of diagrammatic infrared
irregularities is ruled out by general arguments.  Within our
truncation, we can exclude irregularities of non-diagrammatic origin
in the purely transverse subsystem.  Hence it is necessary to solve
the longitudinal system to answer whether the required irregularities
are generated for decoupling-type solutions, which is not done in this
work.  Nevertheless, we are able to produce decoupling-type solutions
by invoking two consistent criteria, which allow for the
differentiation between confining and Higgs-like solutions.  The
decoupling-type solutions are bound by the solution that shows the
minimal mass gap, which is also the solution with minimal back-bending
of the gluon propagator.

\acknowledgments We thank Markus Q. Huber, Axel Maas, Fabian
Rennecke, Andre Sternbeck and Richard Williams for discussions as well
as providing unpublished data.  This work is supported by EMMI, the
grant ERC-AdG-290623, the FWF through Erwin-Schr\"odinger-Stipendium
No.\ J3507-N27, the Studienstiftung des deutschen Volkes, the DFG
through grant STR 1462/1-1, and in part by the Office of Nuclear
Physics in the US Department of Energy's Office of Science 
under Contract No. DE-AC02-05CH11231.

\appendix
\begin{figure*}
	\centering
	\includegraphics[width=0.49\textwidth]{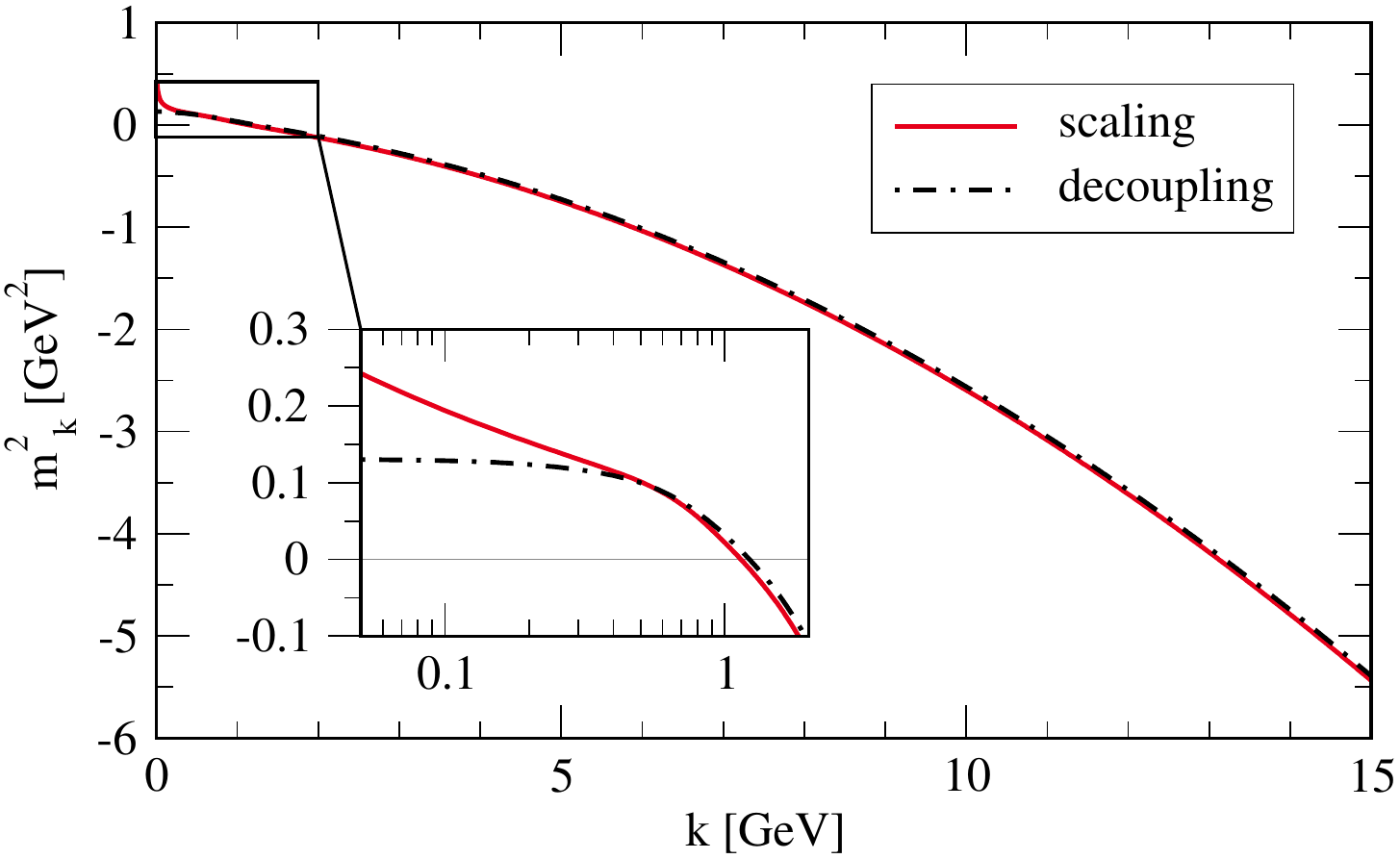}
\hfill
	\includegraphics[width=0.48\textwidth]{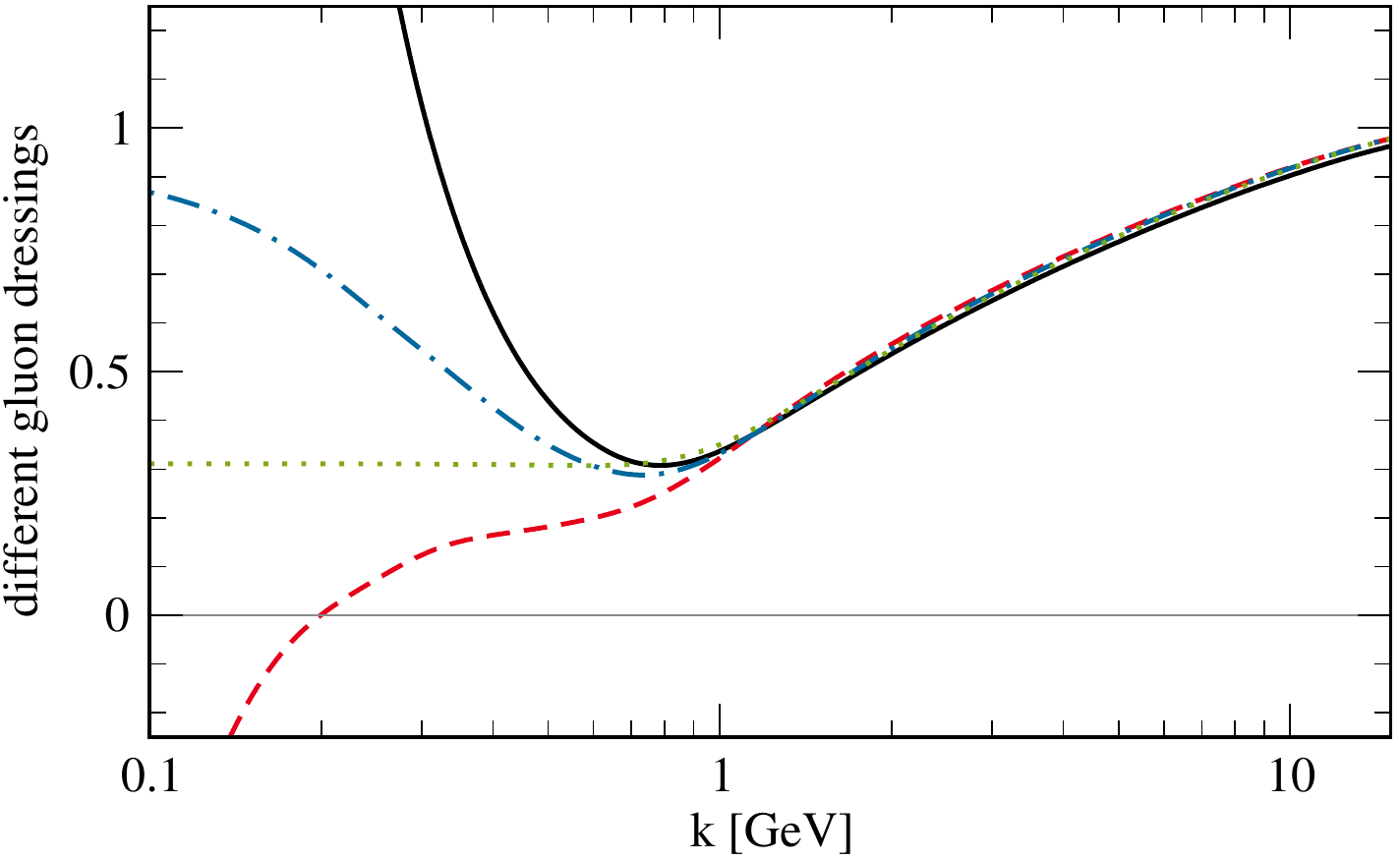}
        \caption{ Left: Gluon mass parameter
          $m^2_k=\Gamma^{(2)}_{AA,k}(p=0)$ over $k\,$.  Right: Possible 
          choices for the scaling prefactor in the gluon regulator:
          $Z_{A,k}(k)$ (black, solid), $\bar{Z}_{A,k}(k)$ (red,
          dashed), $\hat{Z}_{A,k}(k)$ (blue, dot-dashed) and $\tilde
          Z_{A,k}$ (green, dotted) as defined in \eq{eq:reg_dress}
          and \eq{eq:ZAs}.  Independence of the results from the 
          above choice has been checked
          explicitly.  }
	\label{fig:flowQuantities}
\end{figure*}

\section{Gluon mass gap and irregularities}
\label{app:irregularities}
In this section we illustrate the arguments from \Sec{sec:massgap}.
We restrict ourselves to the case of vanishing background fields.  We
first show that the infrared behaviour of the scaling propagators
generically induce a mass gap.  We then demonstrate that the
decoupling solution necessitates irregular vertices for a mass gap
generation due to the infrared finiteness of the decoupling
propagators.  In \App{app:ghosttriangle} we show that the vertex
irregularities required for a decoupling mass gap cannot be of
diagrammatic origin.

A rather general comment is in place here: 
When one is dealing with the gluon mass gap,
it is crucial to carefully take the vanishing 
momentum limit. In the FRG approach this
also means that one must first take the limit
$k\rightarrow 0$ and then $p\rightarrow 0\,$.

\subsection*{Scaling solution}
The infrared-relevant part of the self-energy contribution of the
ghost loop to the inverse gluon propagator is given by
\begin{align}
  \left[\Gamma^{(2),\text{gh-loop}}_{AA}\right]_{\mu\nu}(p) \propto&
  \int_\epsilon^\Lambda \text{d}q \int_{-1}^{1}
  \text{d}t\; q^3\; \sqrt{1-t^2}\nonumber\\[2ex]
  &\quad\cdot \frac{q_\mu}{\left(q^2\right)^{1+\kappa}}
  \frac{(q+p)_\nu}{\left((q+p)^2\right)^{1+\kappa}}\,,
\label{eq:scalingLoop}
\end{align}
where we inserted the infrared ghost propagator from
\eq{eq:scaling_sol} and a classical ghost-gluon vertex, i.e.\
$[\Gamma^{(3)}_{A \bar c c}]_{\mu}^{abc}(p,q) = \imag
f^{abc}q_\mu\,$.  Ignoring the angular integration in
\eq{eq:scalingLoop} for the moment and setting $p=0\,$, we find
\begin{align}
  \left[\Gamma^{(2),\text{gh-loop}}_{AA}\right]_{\mu\nu}(p) \propto
  \int_\epsilon^\Lambda \text{d}q\; q_\mu q_\nu\, q^{-1-4\kappa}\,,
\end{align}
which is infrared-divergent with $\epsilon^{2-4\kappa}$ if
$\kappa>\num{0.5}\,$.  This has to be the case in order to obtain a
divergent gluon mass gap consistent with \eq{eq:scaling_sol}.  To
investigate the mass gap, we project \eq{eq:scalingLoop} with $
\frac{1}{3}\Pi^{\rm \bot}_{\mu\nu}(p)-\Pi^{\rm L}_{\mu\nu}(p)\,$, where
the factor $\frac{1}{3}$ accounts for the three modes of the
transverse projection operator.  We obtain
\begin{align}
  &\left[\Gamma^{(2),\text{gh-loop}}_{AA,\bot}-\Gamma^{(2),\text{gh-loop}}_{AA,L}
  \right](p)\propto\nonumber\\[2ex]
  & \quad \int_0^\Lambda \text{d}q \int_{-1}^{1} \text{d}t\;
  \frac{q^5}{3}\; \sqrt{1-t^2}
  \frac{1-4t^2-\frac{|p|}{|q|}t}{\left(q^2\right)^{1+\kappa}\cdot
    \left((q+p)^2\right)^{1+\kappa}}\,.
\label{eq:scalingMassGap}
\end{align}
One can easily show numerically that the above integral does not vanish in
the limit $p\rightarrow 0$, but diverges with
$\left(p^2\right)^{1-2\kappa}\,$.

\subsection*{Decoupling solution}

Using again the ghost-loop
diagram as an example, we show that a decoupling gluon mass gap requires
irregular vertices.  We choose the ghost-loop diagram since the
ghost-gluon vertex has the smallest tensor space of all vertices,
which makes the example easy to comprehend.  We checked explicitly
that a similar analysis can be carried out for all diagrams and
vertices contributing to the inverse gluon propagator.  This can be
most easily done by assuming regular vertices (which allows to set
$p=0$) and then showing that the mass gap is zero.

To make this point absolutely clear, we demonstrate this argument for
two different tensor bases, for the basis from
\Sec{sec:massgap} and for one with an explicit splitting into 
transverse and longitudinal tensors.  The former basis,
given in \eq{eq:classquantsplit}, reads
\begin{align}
[\Gamma^{(3)}_{A \bar c c}]_{\mu}^{abc}(p,q) = \imag f^{abc}\Bigl(
q_\mu Z_{A \bar cc,\rm cl}(p,q) + p_\mu Z_{A\bar c c,\rm ncl} (p,q)\Bigr) \,,
\label{eq:classquantsplitrepeat}
\end{align}
where $p$ is the gluon and $q$ the anti-ghost momentum.  We assume
that the ghost-gluon vertex is regular. Therefore the second tensor
structure has to be less divergent than $1/|p|$ in the limit of
vanishing gluon momentum, i.e.,
\begin{align}
\label{eq:noGhostGluonVertexPole}
\lim\limits_{|p|\rightarrow 0} |p|\, Z_{A\bar c c,\rm ncl}(q,p) = 0\,.
\end{align}
Note that logarithmic divergences, which for example occur in the
classical tensor structure of the three-gluon vertex and the non-classical
tensor structures of the four-gluon vertex, 
do not suffice to violate their respective equivalents of \eq{eq:noGhostGluonVertexPole}.
Utilising the finiteness of the ghost dressing function and
\eq{eq:noGhostGluonVertexPole}, we can take the limit $|p|\rightarrow
0$ to obtain the mass gap contribution of the ghost loop diagram:
\begin{align}
  &\pat \left(m^2_{\text{gh-loop},\bot}-m^2_\text{gh-loop,L}\right)
  \propto \int_{-1}^{1} \text{d}t\,\sqrt{1-t^2}
  \nonumber\\[2ex]
  &\quad \cdot \left(1-4t^2\right)\, Z_{A \bar cc,\rm cl}(0,|q|,t)\,
  Z_{A \bar cc,\rm cl}(0,|q|,-t)\,\,,
\label{eq:ghostMassGapContr1}
\end{align}
where $\theta=\arccos(t)$ is the angular variable between the loop
momentum and the gluon momentum that is taken to zero.  The dressing
$Z_{A \bar cc,\rm cl}(0,|q|,t)$ is independent of the angular variable
$t$ if the ghost-gluon vertex is regular.  Thus, the mass gap
contribution evaluates to zero:
\begin{align*}
  \int_{-1}^{1} \text{d}t\,\sqrt{1-t^2}\left(1-4t^2\right)=0\,.
\end{align*}
Hence, a gluon mass gap requires requires irregular vertices in the
case of the decoupling solution.
\begin{figure*}
	\centering
	\includegraphics[width=0.48\textwidth]{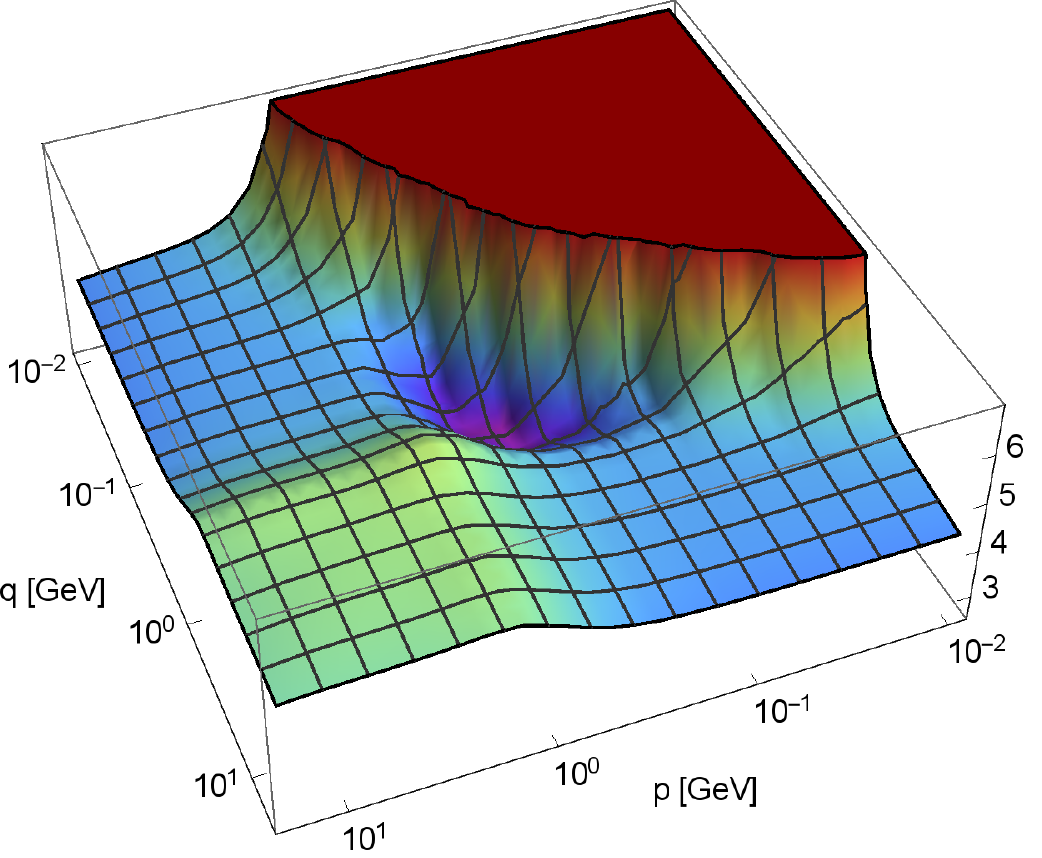}
	\hfill
	\includegraphics[width=0.48\textwidth]{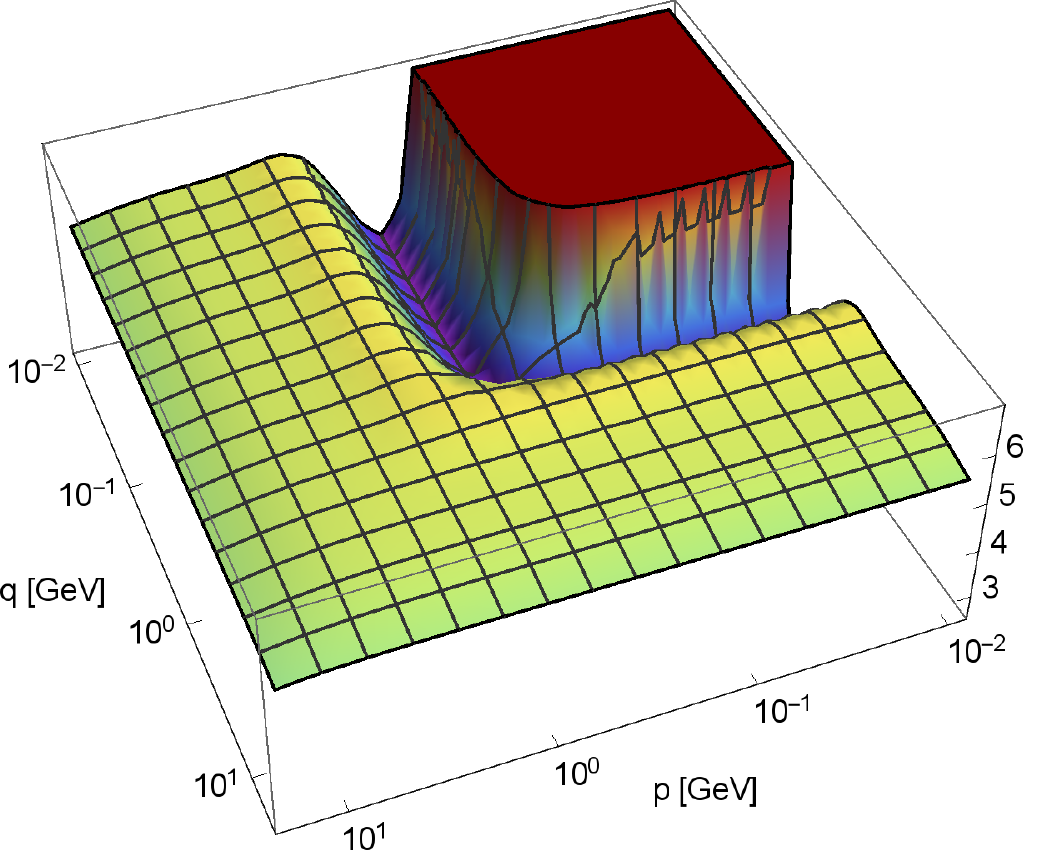}
	\caption{ Left: Four-gluon vertex dressing function
          $Z_{A^4,\bot }(p,\,q,\,0)$ in the tadpole configuration.
          The angular dependence is small compared to the momentum
          dependence. Right: Four-gluon tadpole
          configuration evaluated in the symmetric point
          approximation, showing a quantitative and qualitative
          deviation from the full calculation.}
	\label{fig:fourGluonTadpoleDressing}
\end{figure*}

Since we consider differences between the vanishing longitudinal and
the transverse mass, it might seem more appropriate to split the
tensor basis of the ghost-gluon vertex into a purely longitudinal and
a purely transverse part.  We show now that this leads to the same conclusion.
Transverse and longitudinal projection of the
classical tensor structure already gives us a complete orthogonal
basis:
\begin{align}\nonumber 
  [\Gamma^{(3)}_{A \bar c c}]_{\mu}^{abc}(p,q) = &\imag
  f^{abc}\Biggl[\Pi^{\bot}_{\mu\nu}(p)\, q_\nu Z_{A \bar
    cc,\bot }(p,q) \\[2ex]
  & + \Pi^{\rm L}_{\mu\nu}(p) \,q_\nu Z_{A \bar cc,{\rm L}
  }(p,q)\Biggr] \,,
\label{eq:Lbotsplit}
\end{align}
where $p$ is the gluon and $q$ the anti-ghost momentum. The projection
operators are given by $\Pi^{\rm L}_{\mu\nu}(p) =p_\mu p_\nu/p^2$ and
$\Pi^{\bot}_{\mu\nu}(p) = \id_{\mu\nu} - \Pi^{\rm L}_{\mu\nu}(p)$.
Note that the basis \eq{eq:Lbotsplit} contains a discontinuity at
$p=0$ due to the projection operators. The mass gap contribution of
the ghost diagram with this ghost-gluon vertex basis evaluates to
\begin{align}\nonumber 
		&\pat \left(m^2_{\text{gh-loop},\bot}-m^2_\text{gh-loop,L} \right)
		\propto \int_{-1}^{1} \text{d}t\,\sqrt{1-t^2}
		\\[2ex]\nonumber 
		&\quad \cdot\Big(\frac{1-t^2}{3}Z_{A \bar
			cc,\bot}(0,|q|,t)\,Z_{A \bar
			cc,\bot}(0,|q|,-t)\\[2ex]
		&\quad\quad\quad-t^2\, Z_{A \bar cc,{\rm
				L} }(0,|q|,t)\,Z_{A \bar cc,{\rm
				L} }(0,|q|,-t)\Big)\,\,.
	\label{eq:ghostMassGapContr2}
\end{align}
Regularity \eq{eq:noGhostGluonVertexPole}, implies a degenerate 
tensor space in the limit of vanishing gluon
momentum.  The ghost-gluon vertex can then be fully described by $Z_{A
  \bar cc,\rm cl}(0,|q|)\equiv Z_{A \bar cc,{\rm L} }(0,|q|,t)$.  Using the identity $\id_{\mu\nu} =
\Pi^{\bot}_{\mu\nu}(p) + \Pi^{\rm L}_{\mu\nu}(p)$, we find
\begin{align} 
Z_{A\bar cc,\rm cl}(0,|q|) = Z_{A \bar cc,\bot}(0,|q|) = Z_{A \bar cc,{\rm L} }(0,|q|)\,.
\label{eq:degenerateTensorSpace}
\end{align}
Using \eq{eq:degenerateTensorSpace} we can perform the angular integration in 
\eq{eq:ghostMassGapContr2} and find that the mass gap contribution vanishes.

We want to stress that this statement is general and holds for any
diagrammatic method.  For example, the same conclusion can be drawn
for the ghost-loop diagram of the gluon propagator Dyson-Schwinger
equation that is also proportional to \eq{eq:ghostMassGapContr1} or
\eq{eq:ghostMassGapContr2}. Consequently, for the decoupling solution
there can be no mass gap with regular vertices.

\begin{figure*}
	\centering
	\includegraphics[width=0.48\textwidth]{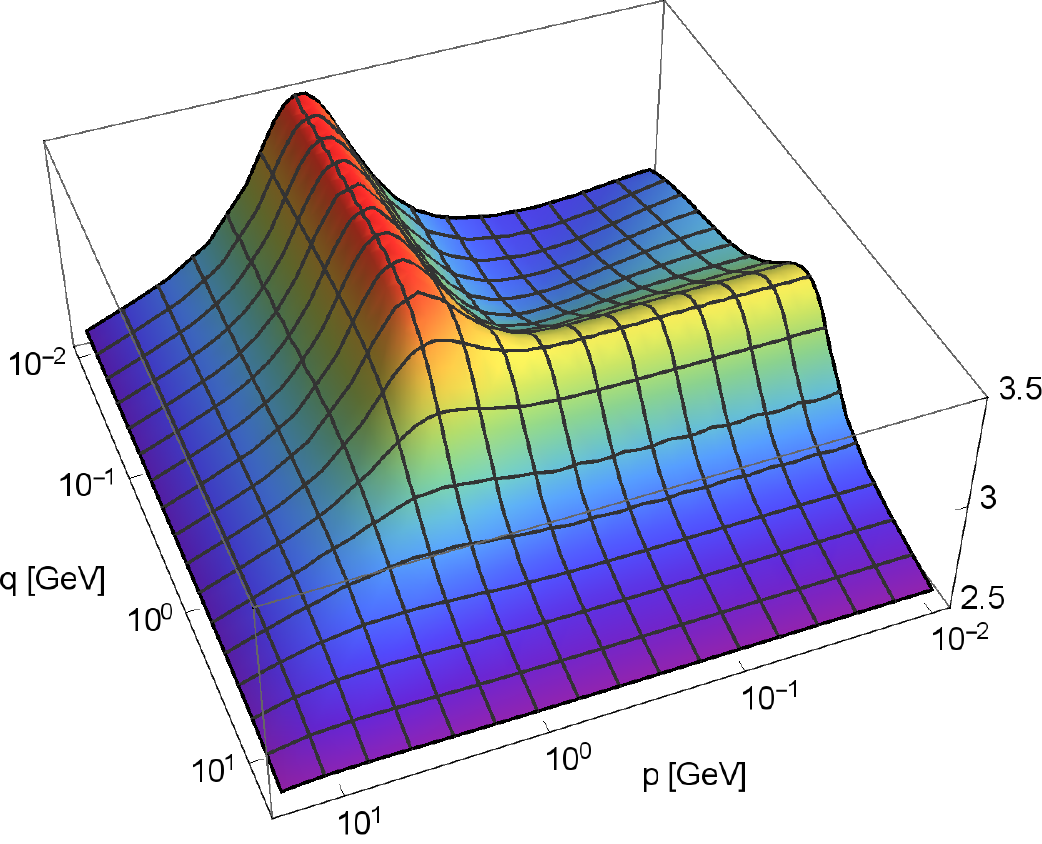}
	\hfill
	\includegraphics[width=0.48\textwidth]{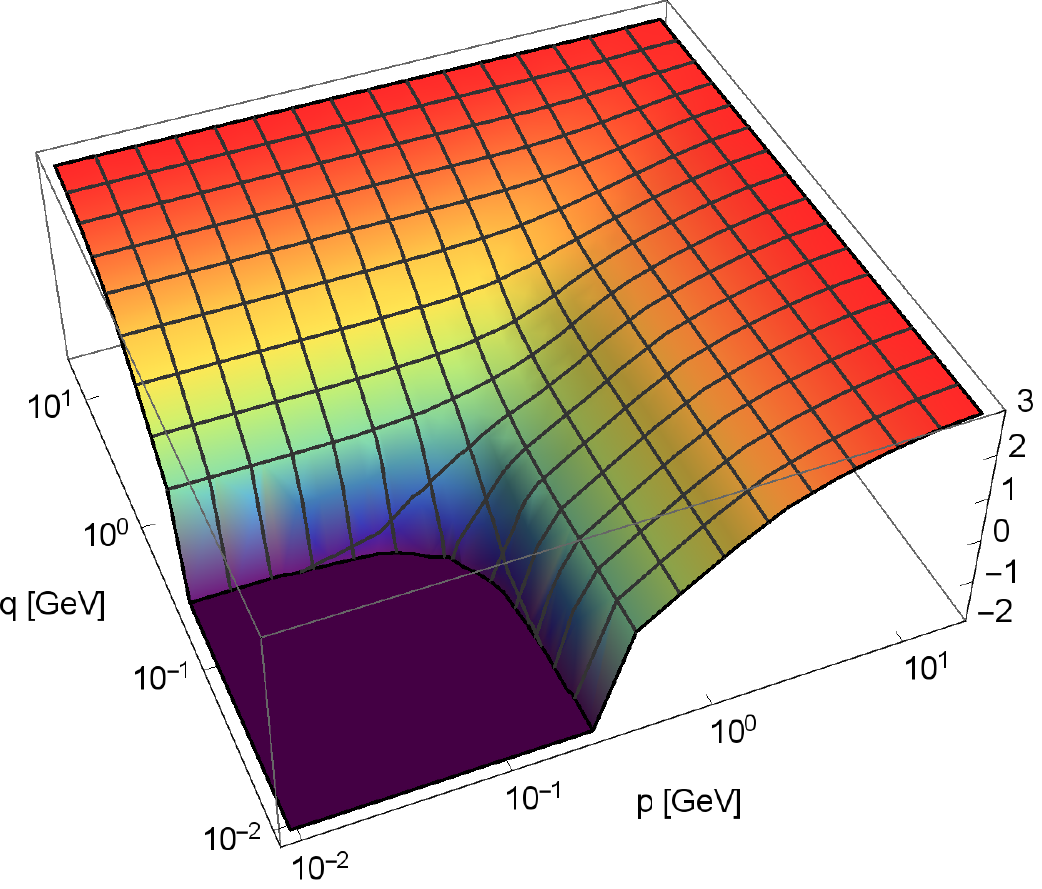}
	\caption{
		Left: Ghost-gluon vertex dressing function $Z_{A \bar cc,\bot }(p,\,q,\,0)\,$.
		Right: Three-gluon vertex dressing function $Z_{A^3,\bot }(p,\,q,\,0)\,$.
	}
	\label{fig:ghostGluonDressing3D_threeGluonDressing3D}
\end{figure*}

\section{Ghost-triangle}
\label{app:ghosttriangle}
In \Sec{app:irregularities} it is shown that the decoupling solution 
requires irregular vertices.
In the three gluon vertex, this irregularity has to occur if 
one momentum is sent to zero while the others are non-vanishing.
Those vertex irregularities can be generated either 
by back-coupling of momentum dependence or by 
diagrammatic infrared singularities.
We cannot observe the former in our computation 
of the purely transverse system.
To investigate the latter case we use that the 
gluonic diagrams decouple from 
the infrared dynamics due to the gluon mass gap.
Therefore we can focus on the 
ghost loops as possible sources of diagrammatic
 IR irregularities without loss of generality.
The ghost-gluon vertex as well as the ghost propagator are constant and finite
in the infrared.
In the following we show explicitly that the three-gluon vertex
does not obtain an irregular contribution from the ghost triangle.
Its relevant part is given by
\begin{align}
	\label{eq:ghostTriangle}
	[\Gamma_{A^3}^{(3),\text{gh-loop}}]_{\mu\nu\rho}(p,q,r)\propto\int
    \frac{\text{d}^\text{d}l}{(2 \pi)^\text{d}}
    \frac{(l+p)_\mu}{(l+p)^2} \frac{l_\nu}{l^2}
    \frac{(l-q)_\rho}{(l-q)^2}\,.
\end{align}
To confirm that \eq{eq:ghostTriangle} does not generate an
irregularity in the limit $\frac{|p|}{|q|}\rightarrow 0$, we consider
the low and high momentum integration regions separately.  If the
loop momentum $|l|$ is of the order of $|q|$, then $|p|\ll|l|$ and
the $p$ dependence in \eq{eq:ghostTriangle} is suppressed.  Thus no
irregular structure can be generated from this integration region.
For small loop momenta $|l|\approx |p|$ we have $|l|\ll|q|$ and the
contribution to the integral in the limit $\frac{|p|}{|l|}\rightarrow
0$ is given by
\begin{align}
\label{eq:reducedGhostTriangle}
	\frac{q_\rho}{q^2} \int \frac{\text{d}^\text{d}l}{(2 \pi)^\text{d}}
	\frac{(l+p)_\mu}{(l+p)^2}
	\frac{l_\nu}{l^2}\,.
\end{align}
This integral can easily be solved analytically for $d=4-2\epsilon$
to show that it has no irregularities, which one also expects from a
dimensional analysis of \eq{eq:reducedGhostTriangle}.  Hence, we
conclude, that the decoupling ghost triangle cannot generate the
irregularity necessary for the dynamical generation of a gluon mass
gap. Note that the ghost triangle develops a non-trivial pole
structure in the case of the scaling solution, see
\cite{Alkofer:2008dt}.  We have verified these findings numerically,
and since they are in accordance with perturbation theory, we expect
similar arguments to hold for the ghost loops contributing to higher
$n$-point functions.

\section{Numerical implementation}
\label{app:technicalDetails}
The algebraic flow equations are derived using
DoFun~\cite{Huber:2011qr}.  The projected flow equations are then
traced using the \textit{FormTracer}~\cite{formtracer}, a Mathematica
package that uses FORM
\cite{Vermaseren:2000nd,vanRitbergen:1998pn}.  The output is exported
as optimized C code with FORMs optimisation algorithm
\cite{Kuipers:2013pba}.  The calculation is performed with the
\textit{frgsolver}, a flexible, object-orientated, parallelised C++ framework
developed by the fQCD collaboration \cite{fQCD}, whose development was
initiated in \cite{Mitter:2014wpa}.  The framework uses the adaptive
ordinary differential equation solver from the BOOST libraries
\cite{odeboost}, the Eigen linear algebra library \cite{eigen} and an
adaptive multidimensional integration routine from \cite{cubature}
which implements \cite{berntsen1991adaptive,genz1980remarks} to solve
the integro-differential equations.

\begin{figure*}
  \centering \subfloat[$Z_{A \bar cc,\bot
  }(p,\,p,\,-\frac{1}{2})\,$.]{\includegraphics[width=0.49\textwidth]{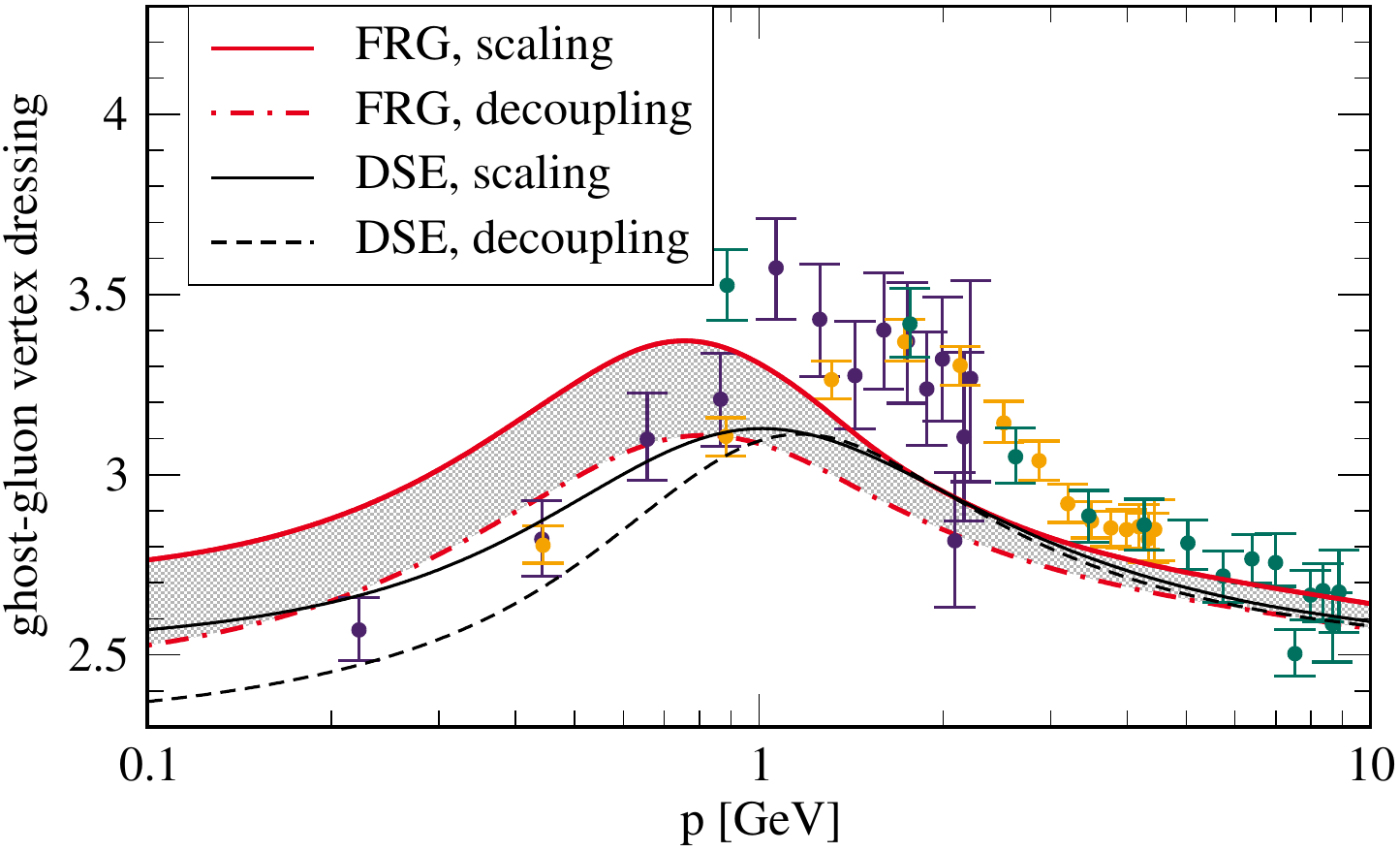}}
  \hfill \subfloat[$Z_{A^3,\bot
  }(p,\,p,\,-\frac{1}{2})\,$.]{\includegraphics[width=0.48\textwidth]{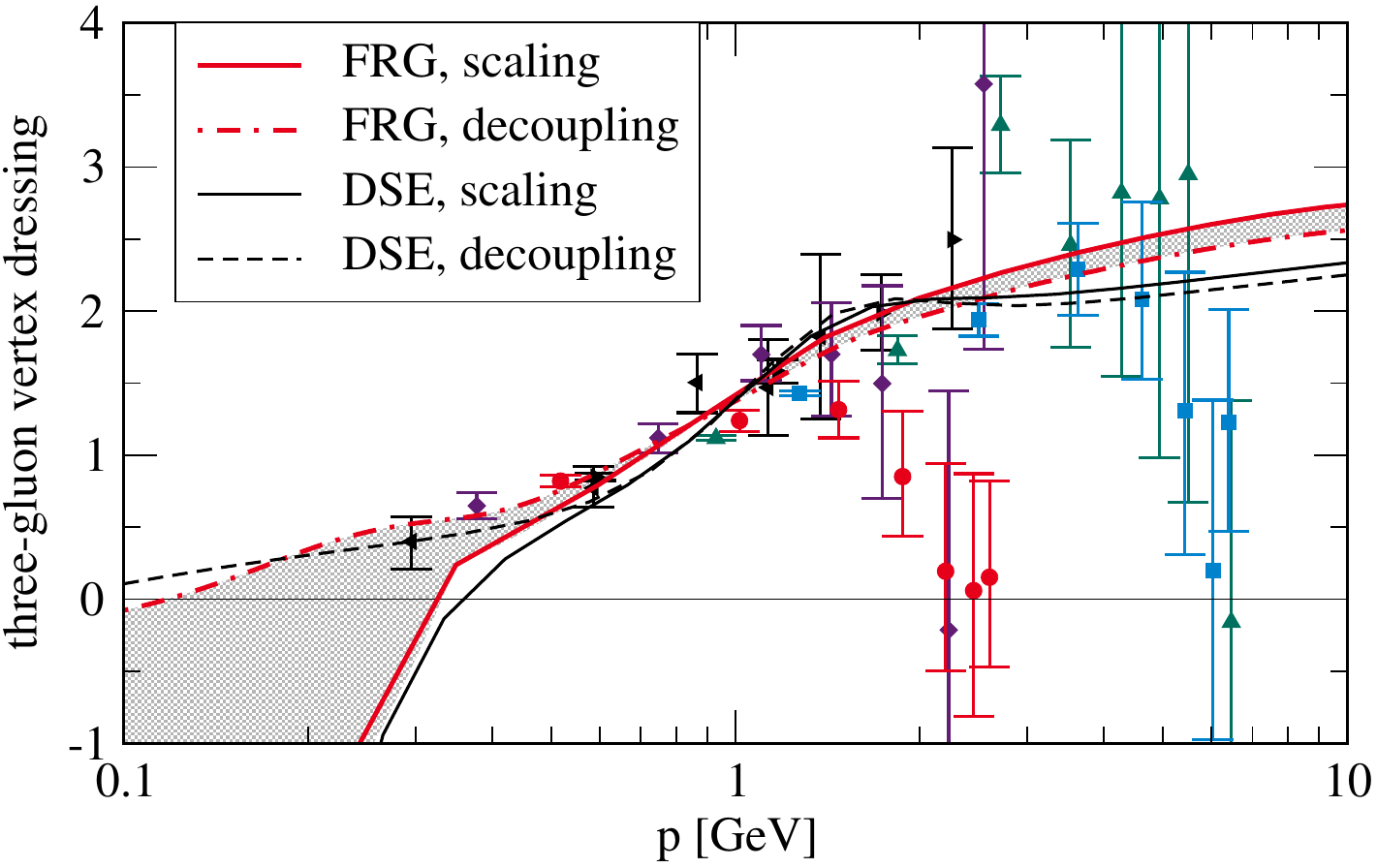}}
  \\
  \subfloat[$Z_{A \bar cc,\bot
  }(p,\,p,\,0)\,$.]{\includegraphics[width=0.49\textwidth]{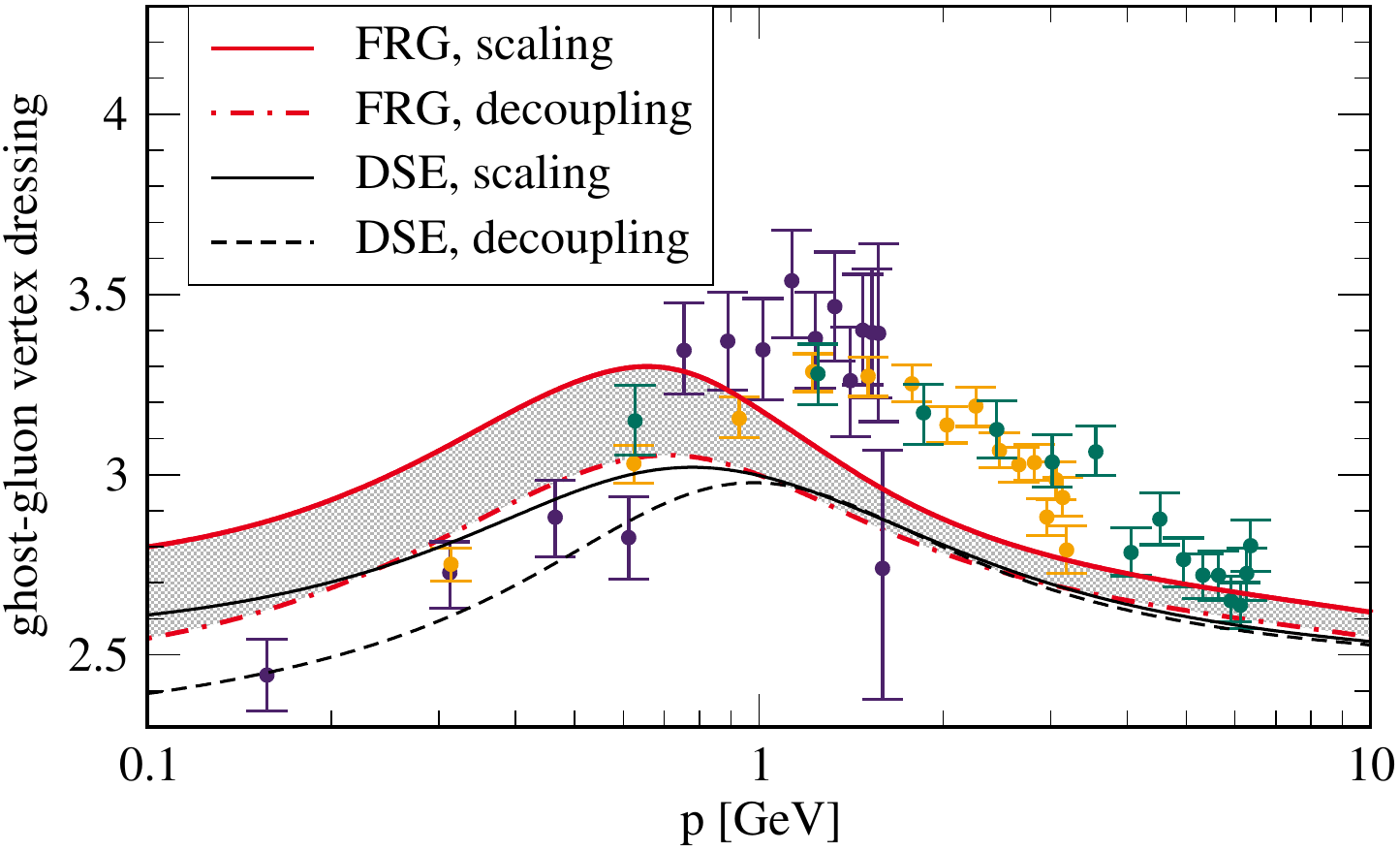}}
  \hfill \subfloat[$Z_{A^3,\bot
  }(p,\,p,\,0)\,$.]{\includegraphics[width=0.48\textwidth]{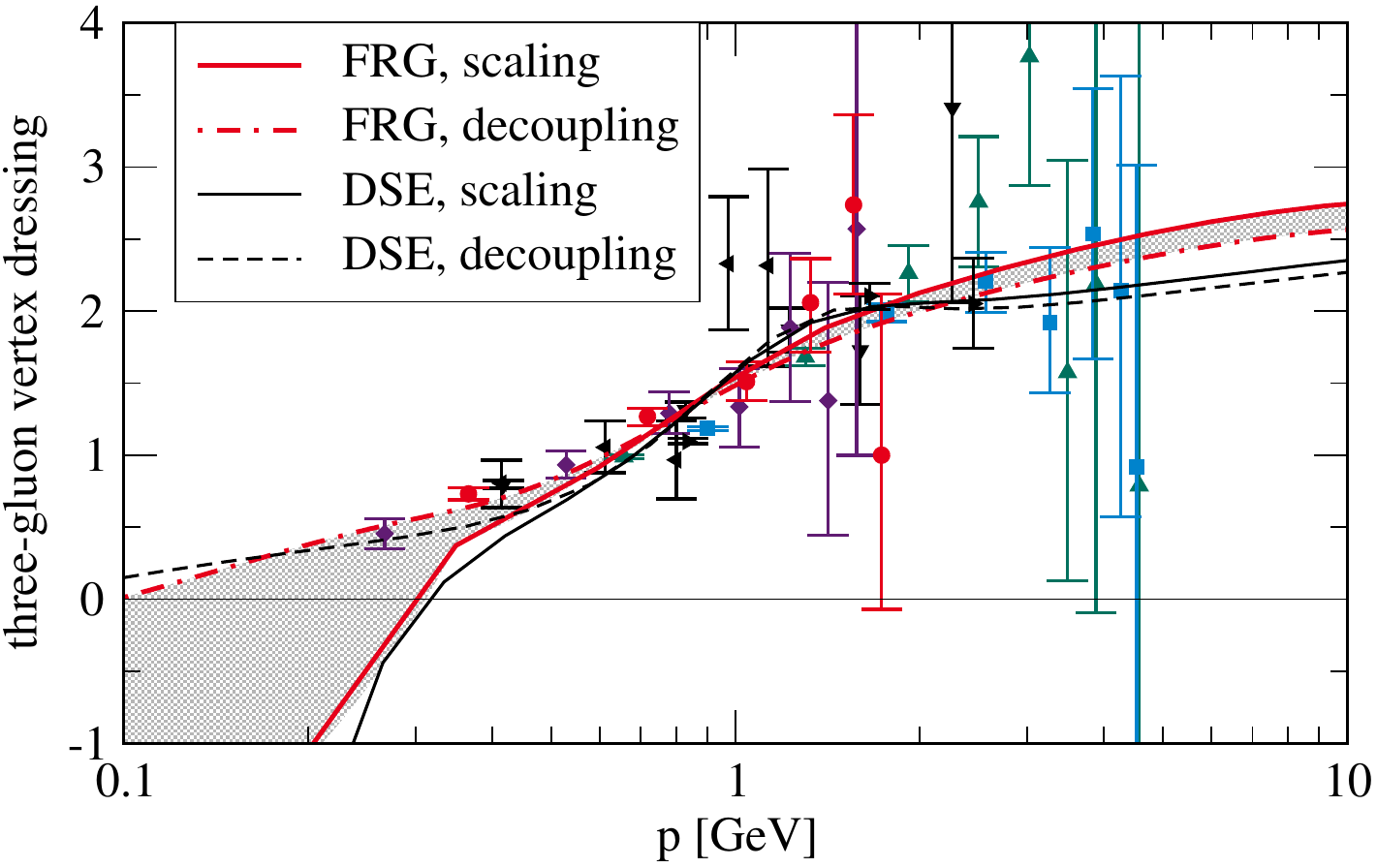}}
  \\
  \subfloat[$Z_{A \bar cc,\bot
  }(p,\,0,\,0)\,$.]{\includegraphics[width=0.49\textwidth]{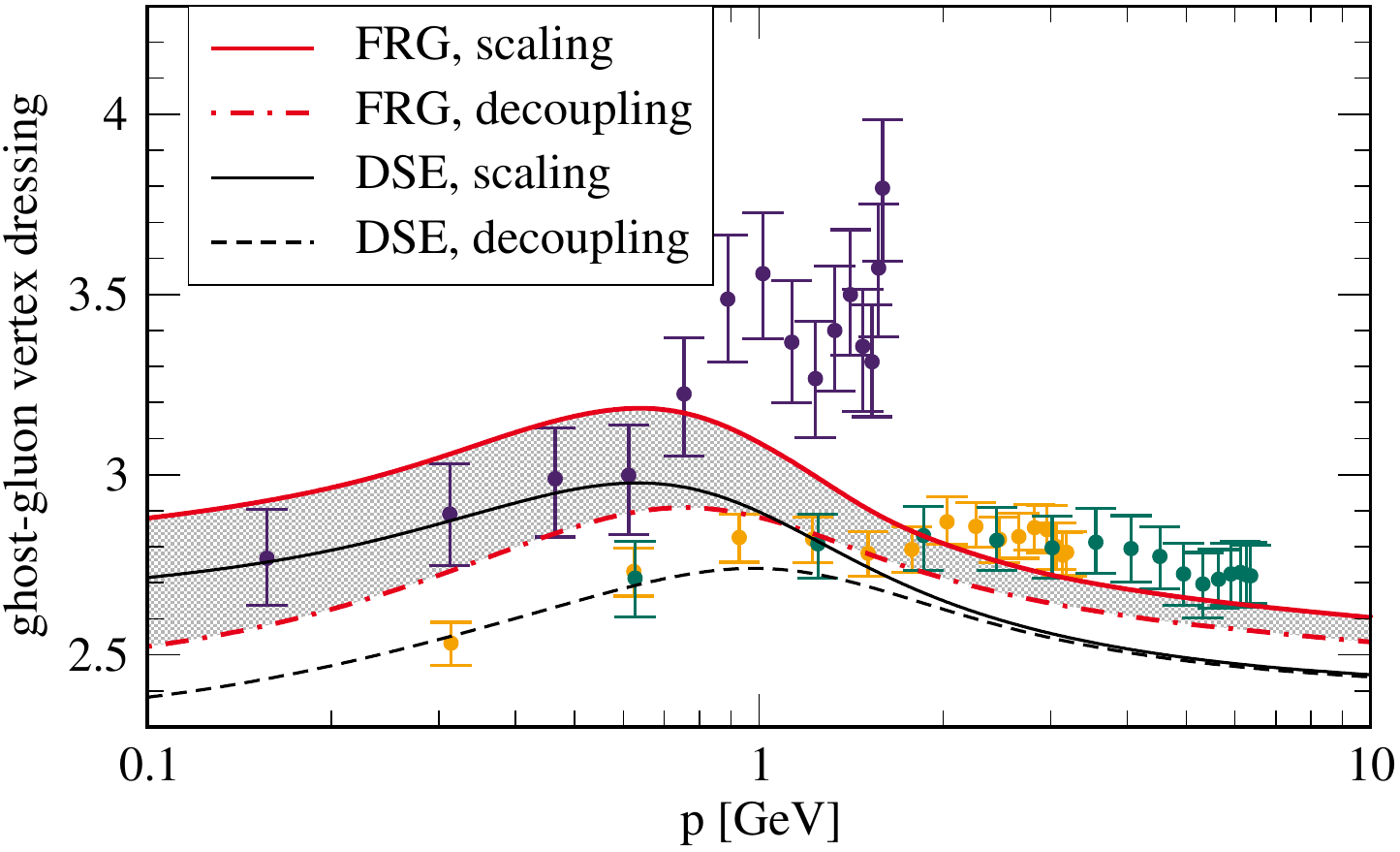}}
  \hfill \subfloat[$Z_{A^3,\bot
  }(p,\,0,\,0)\,$.]{\includegraphics[width=0.48\textwidth]{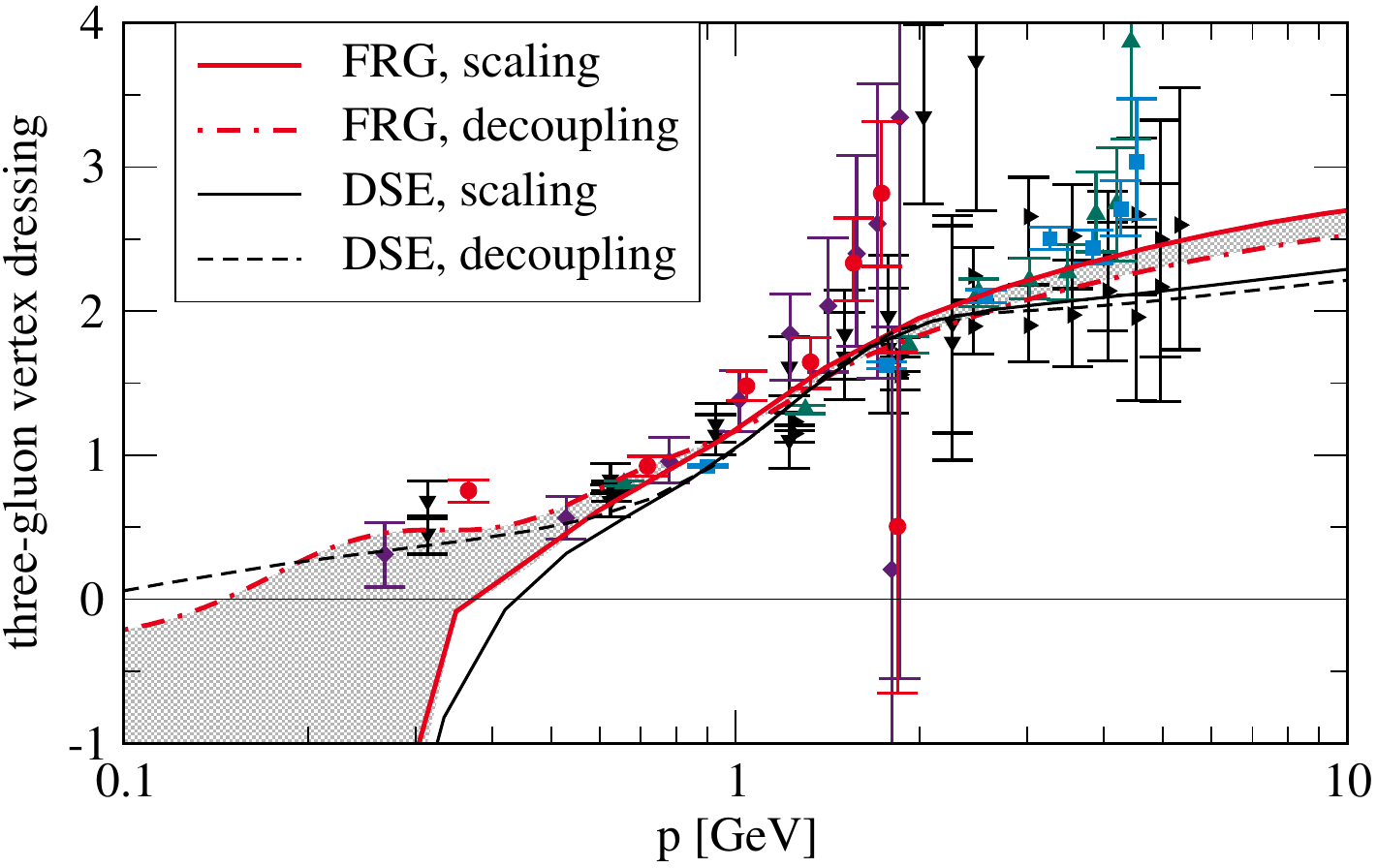}}

        \caption{ Left: Ghost-gluon vertex dressing function $Z_{A
            \bar cc,\bot }(p,\,q,\,\cos\sphericalangle (p,\,q))$ in
          comparison to $SU(2)$
          lattice~\cite{Cucchieri:2006tf,Cucchieri:2008qm,Maas:preparation}
          and DSE results~\cite{Huber:2012kd,Huber:unpublished}.  The
          lattice results are obtained from $N=32^4$ lattices.  The
          magenta/orange/green points (colour online) correspond to
          $\beta \in\{2.13,\,2.39,\,2.60\}$ and lattice spacing
          $a^{-1}
          \in\{\SI{0.8}{\GeV},\,\SI{1.6}{\GeV},\,\SI{3.2}{\GeV}\}\,$,
          respectively.
          \\
          Right: Three-gluon vertex dressing function $Z_{A^3,\bot
          }(p,\,q,\,\cos\sphericalangle (p,\,q))$ compared with
          $SU(2)$
          lattice~\cite{Cucchieri:2006tf,Cucchieri:2008qm,Maas:preparation}
          and Dyson-Schwinger~\cite{Blum:2014gna} results.  The
          coloured lattice points are taken from
          \cite{Cucchieri:2006tf,Cucchieri:2008qm} and correspond to
          $\beta\in \{2.2, 2.5\}$ and different lattice sizes $1.4
          \text{ fm}<L<4.7 \text{ fm}$.  The lattice results shown in
          black are based on \cite{Cucchieri:2006tf,Cucchieri:2008qm}
          but stem from \cite{Maas:preparation}.  These are gained
          from $N \in \{24^4,\,32^4\}$ lattices with $\beta
          \in\{2.13,\,2.39,\,2.60\}$ and lattice spacing $a^{-1}
          \in\{\SI{0.8}{\GeV},\,\SI{1.6}{\GeV},\,\SI{3.2}{\GeV}\}\,.$
          \\
          The comparison with $SU(2)$ lattice simulations is justified
          since the propagators of $SU(2)$ and $SU(3)$ Yang-Mills
          theory agree well for a large range of momenta
          \cite{Cucchieri:2007zm,Sternbeck:2007ug} after a respective
          normalisation in this regime.  We rescaled all DSE results
          such that they match the scaling solution in the symmetric
          point configuration at $p=\SI{2}{\GeV}\,$.  Note that the
          scaling and decoupling solutions differ in the UV just due
          to the different field renormalisations, cf.\
          \Fig{fig:main_result}.  The physically relevant couplings,
          given by \eq{eq:runcoup}, agree in the UV.\\}
	\label{fig:ThreePointVertices}
\end{figure*}
\section{Tensor structures of YM-vertices}
\label{app:tensorstructures}
In this section we define our conventions for the tensor structures in
which we expanded our vertices and the used projections.  The tensor
structures for the classical YM three-point vertices are defined by
\begin{align}
\left[\mathcal{T}_{A^3,{\rm cl}}\right]^{abc}_{\mu\nu\rho}(p,q) &= \imag f^{abc}
\Big\{(p-q)_\rho
\delta_{\mu\nu}+ \text{ perm.}\Big\} \,, 
\nonumber\\[2ex]
\left[\mathcal{T}_{A \bar c c,{\rm cl}}\right]^{abc}_{\mu}(p,q) &=
\imag f^{abc} q_\mu  \,,
\end{align}
and by
\begin{align}
\left[\mathcal{T}_{A^4,{\rm cl}}\right]^{abcd}_{\mu\nu\rho\sigma}(p,q,r) &=
f^{abn}f^{cdn}\delta_{\mu\rho}\delta_{\nu\sigma} + \text{
	perm.}\,,
\end{align}
for the four-point function. For the transversally projected
ghost-gluon vertex this single tensor constitutes already a full basis
and the projection is uniquely defined. However, additional allowed
tensors exist in the case of the three-gluon and four-gluon vertices.
We obtained the dressing functions by contracting the equations with
\begin{align}
\Pi^{\bot}_{\mu \bar
	\mu}(p)\Pi^{\bot}_{\nu\bar\nu}(q)\Pi^{\bot}_{\rho\bar\rho}(p+q)
\left[\mathcal{T}_{A^3,{\rm cl}}\right]^{abc}_{\bar\mu\bar\nu\bar\rho}(p,q)\,,
\end{align}
and
\begin{align}
\Pi^{\bot}_{\mu \bar
	\mu}(p)\Pi^{\bot}_{\nu\bar\nu}(q)\Pi^{\bot}_{\rho\bar\rho}(r)
\Pi^{\bot}_{\sigma\bar\sigma}(p+q+r)
\left[\mathcal{T}_{A^4,{\rm cl}}\right]^{abcd}_{\bar\mu\bar\nu\bar\rho\bar\sigma}(p,q,r)\,,
\end{align}
respectively.

\section{Regulators}
\label{app:regulator}

In the functional renormalisation group, the choice of the regulator,
together with the choice of the cutoff-independent parts of
the initial effective action corresponds to defining a
renormalisation scheme, for a more detailed discussion see
\cite{Pawlowski:2005xe}. Moreover, to any given order of a given
approximation scheme there exist optimised regulators that lead to
the most rapid convergence of the results, hence minimising the
systematic error, see
\cite{Litim:2000ci,Litim:2001up,Pawlowski:2005xe}. For recent
extensions and applications relevant for the present work see
\cite{Pawlowski:2015mlf,Schnoerr:2013bk}.  In the present work we
use
\begin{align}\nonumber 
          R^{ab}_{k,\mu\nu}(p) &= \tilde Z_{A,k}\, r(p^2/k^2)\, p^2\,
          \delta^{ab}\, \Pi^{\bot}_{\mu\nu}(p) \,,\\[2ex]
          R^{ab}_k(p) &= \tilde Z_{c,k}\, r(p^2/k^2)\, p^2
          \delta^{ab}\,,
          \label{eq:regulators}
\end{align}
for the gluon and the ghost fields, respectively.  For the shape
function we choose a smooth version of the Litim or flat regulator
\cite{Litim:2000ci}:
\begin{align}
	r(x) &= \left(\frac{1}{x}-1\right) \cdot \frac{1}{1+e^{\frac{x-1}{a}}}\,, 
\end{align}
where we set $a=0.02\,$. It has been argued in
\cite{Pawlowski:2005xe} that smooth versions of the flat regulator
satisfy the functional optimisation criterion put forward there.

In \eq{eq:regulators} we multiply the regulators with scaling factors 
$\tilde Z\,$, related to the corresponding wave function renormalisations of the
gluon and ghost fields
\begin{align}
  \tilde Z_{A,k} &\coloneqq Z_{A,k}((k^n+\bar{k}^n)^{1/n})\ ,\nonumber  \\[2ex]
  \tilde Z_{c,k} &\coloneqq Z_{c,k}(k)\,,
  \label{eq:reg_dress}
\end{align}
where we choose $n\approx6$ and $\bar k \approx \SI{1}{\GeV}\,$. The
cutoff scale running of $\tilde Z_A\,$ is held constant below scales of about
$\SI{1}{\GeV}$ as the gluon wave function renormalisation $Z_{A,k}(p\approx k)\,$
diverges for $k\to 0$. Separating the tensor structure by
\begin{align}
  [\Gamma^{(2)}_{AA}]^{ab}_{\mu\nu}(p) &\eqqcolon
  \Gamma^{(2)}_{AA,k}(p) \, \delta^{ab}\, \Pi^{\bot}_{\mu\nu}(p)\,,
\end{align}
we parameterise $\Gamma^{(2)}_{AA,k}(p)$ by
\begin{align}
  \nonumber \Gamma^{(2)}_{AA,k}(p) &\eqqcolon Z_{A,k}(p) \cdot p^2
  \\[2ex]\nonumber
  &\eqqcolon \bar{Z}_{A,k}(p)\cdot p^2 + m^2_k\\[2ex]
  &\eqqcolon \hat{Z}_{A,k}(p)\cdot(p^2+m^2_k) \,,
 \label{eq:ZAs}
 \end{align}
 where we define $m^2_k:=\Gamma^{(2)}_{AA,k}(0)$ to guarantee the
 uniqueness of $\hat{Z}_{A,k}$.  We see that these choices differ
 considerably below $\SI{1}{\GeV}$.  For more details see
 \Fig{fig:flowQuantities} (right panel). In particular the naive choice $Z_{A,k}$
 diverges since it carries the gluon mass gap.  Consequently, we
 freeze $Z_{A,k}$ at a scale $\bar k$ close to $\SI{1}{\GeV}$.  We have
 checked explicitly that varying the value of $\bar k$ and $n$ has no
 influence on our results.

\section{Scale setting and normalisation}
\label{app:rescaling}

When comparing to lattice results, the momentum scales as well as the
global normalisations of the fields have to be fixed. We
set the scale by
\begin{align*}
 p_{\si{GeV}}=c \cdot p_\text{internal}\,,
\end{align*}
where we choose $c$ such that the scale of the maximum of the gluon
dressing $1/Z_A(p)\,$ agrees with the lattice scale from
\cite{Sternbeck:2006cg}, which lies at $p\approx\SI{0.955}{\GeV}\,$.

We then rescale the gluon dressing by $Z_A^{-1}(p)\to a\; Z_A^{-1}(p)$
with $a$ chosen such that it minimises
\begin{align}
  &N_{Z_A}(a)=\sum_{i} \frac{\Delta x_i}{\Delta E_i^2}\cdot\Big[
  \left(a\, Z_A^{-1}(p_i)-Z^{L,-1}_A(p_i)\right)^2
  \nonumber\\
  &\quad+ \left(a\, \partial_p Z_A^{-1}(p_i)- \partial_p
    Z^{L,-1}_A(p_i)\right)^2 \Big]\,,
	\label{eq:Znorm}
\end{align}
where we sum over all lattice points that fulfil $\SI{0.8}{\GeV}\leq
p_i\leq \SI{4}{\GeV}\,$.  We do not include points with smaller
momenta since they can be affected by the global gauge fixing
procedure.  Points with momentum larger than $\SI{4}{\GeV}$ are also
not included since they might contain finite volume effects.  In
\eq{eq:Znorm}, we weight the lattice points with $\Delta x_i / \Delta
E_i^2\,$, where $\Delta x_i$ denotes the distance to the next point
and $\Delta E_i$ is the statistical error of the point.  The
superscript $L$ in \eq{eq:Znorm} marks lattice dressing functions.
The ghost dressing is rescaled analogously.

\bibliography{../../../bib_master}

\end{document}